\documentclass[prd,nofootinbib,showpacs]{revtex4}
\usepackage{graphics,color}
\usepackage{bm}
\usepackage{graphicx}
\usepackage{amsmath}
\usepackage{amssymb}
\usepackage{enumitem}
\usepackage{tensor}	
\usepackage[fleqn]{mathtools}
\usepackage{epstopdf}

\def\be{\begin{equation}}
\def\ee{\end{equation}}
\def\ba{\begin{eqnarray}}
\def\ea{\end{eqnarray}}
\def\l{\left}
\def\r{\right}
\def\f{\frac}

\def\hub{{\mathcal H}}
\def\zed{{\mathcal Z}}
\newcommand{\SDer}{\bar{\nabla}}

\begin{document}

\title{Effective Field Theory of Cosmic Acceleration:\\
 an implementation  in CAMB}

\author{Bin Hu$^{1}$, Marco Raveri$^{2,3}$, Noemi Frusciante$^{2,3}$, Alessandra Silvestri$^{2,3,4}$}

\smallskip
\affiliation{$^{1}$ Institute Lorentz, Leiden University, PO Box 9506, Leiden 2300 RA, The Netherlands \\
\smallskip 
$^{2}$ SISSA - International School for Advanced Studies, Via Bonomea 265, 34136, Trieste, Italy \\
\smallskip
$^{3}$ INFN, Sezione di Trieste, Via Valerio 2, I-34127 Trieste, Italy \\
\smallskip
$^{4}$ INAF-Osservatorio Astronomico di Trieste, Via G.B. Tiepolo 11, I-34131 Trieste, Italy}

\begin{abstract}
We implement the effective field theory (EFT) approach to dark energy and modified gravity in the public Einstein-Boltzmann solver CAMB. The resulting code, which we dub EFTCAMB,  is a powerful and versatile tool that can be used for several objectives. It can be employed to evolve the full dynamics of linear scalar perturbations of a broad range of single field dark energy and modified gravity model, once the model of interest is mapped into the EFT formalism. It offers a numerical implementation of EFT as a model-independent framework to test gravity on cosmological scales. EFTCAMB has a built-in check for the fulfillment of general stability conditions such as the absence of ghost and superluminal propagation of perturbations. It handles phantom-divide crossing models and does not contain any quasi-static approximation, but rather evolves the full dynamics of perturbations on linear scales. As we will show, the latter is an important feature in view of the accuracy and scale 
range of upcoming surveys.
We show the reliability and applicability of our code by evolving the dynamics of linear perturbations and extracting predictions for power spectra in several models. In particular we perform a thorough analysis of $f(R)$ theories, comparing our outputs with those of an existing code for $\Lambda$CDM backgrounds, and finding an agreement that can reach $0.1\%$ for models with a Compton wavelength consistent with current cosmological data.
We then showcase the flexibility of our code studying two different scenarios. First we produce new results for designer $f(R)$ models with a time-varying dark energy equation of state. Second, we extract predictions for linear observables in some parametrized EFT models with a phantom-divide crossing equation of state for dark energy.
\end{abstract}

{ 
\maketitle
\section{Introduction}\label{Intro}
An outstanding problem faced by  modern cosmology is cosmic acceleration, i.e. the phase of accelerated expansion recently entered by the Universe~\cite{Riess:1998cb, Perlmutter:1998np}, for which we still lack a satisfactory theoretical explanation. Within the context of  General Relativity (GR), an accelerated expansion can be achieved adding an extra ingredient in the energy budget of the Universe, commonly referred to as dark energy (DE). The latter can be static, as in the standard cosmological model ($\Lambda$CDM), or dynamical.  Alternatively, one can conceive models that modify the laws of gravity on large scales in order to achieve self-accelerating solutions in the presence of negligible matter. We will refer to the latter with the general term `modified gravity' (MG). A plethora of models addressing the phenomenon of cosmic acceleration have been proposed and analyzed in the past fifteen 
years~\cite{Copeland:2006wr,Silvestri:2009hh,Clifton:2011jh,Amendolabook}, and it has become increasingly evident that the dynamics of perturbations  will offer precious information to  discern among candidate models, breaking, at least partially, the degeneracy that characterize them at the background level~\cite{Lue:2004rj,Brax:2004qh,Lazkoz:2006gp,Song:2006ej,Bean:2006up,Dvali:2007kt,Appleby:2007vb,Pogosian:2007sw,Brax:2008hh}.

Anticipating a wealth of high precision large scale structure data from ongoing and upcoming surveys, such as \emph{Planck}~\cite{planck}, SDSS~\cite{sdss}, DES~\cite{des}, LSST~\cite{lsst} and Euclid~\cite{euclid},  it is important to identify a model- independent way of testing the theory of gravity against the evolution of linear cosmological perturbations. To this extent, several proposals have been put forward and analyzed in the past years~\cite{Bertschinger:2006aw,Linder:2007hg,Amendola:2007rr,Zhang:2007nk,Hu:2007pj,Daniel:2008et,Song:2008vm,Skordis:2008vt,Song:2008xd,Zhao:2009fn,Zhao:2010dz,Song:2010rm,Daniel:2010ky,Pogosian:2010tj,Bean:2010zq,Thomas:2011pj,Baker:2011jy,Thomas:2011sf,Zhao:2011te,Hojjati:2011xd,Bertschinger:2011kk,Brax:2011aw,Brax:2012gr,Sawicki:2012re,Baker:2012zs,Amendola:2012ky,Hojjati:2012ci,Motta:2013cwa,Asaba:2013xql,Baker:2013hia,Dossett:2013npa}. 
In this paper we focus on a recent proposal which applies the effective field theory (EFT) formalism to the phenomenon of cosmic acceleration~\cite{Gubitosi:2012hu,Bloomfield:2012ff,Piazza:2013coa}; see also~\cite{Creminelli:2008wc,Park:2010cw,Battye:2012eu,Cheung:2007st,Weinberg:2008hq,Jimenez:2011nn,Carrasco:2012cv,Hertzberg:2012qn} for  previous work in the context of inflation, large scale structure and quintessence. The formalism is based on an action constructed in unitary gauge out of all operators that are consistent with time-dependent spatial diffeomorphisms and are ordered according to the power of perturbations and derivatives. At each order in perturbations, there is a finite number of such operators that enter the action multiplied by  time-dependent coefficients to which we will refer as EFT functions. The background dynamics is determined solely by three EFT functions, that are the coefficients of the three background operators; while the general dynamics of linear scalar perturbations is affected by a total of six operators, and can therefore be analyzed 
in terms of six time-dependent functions. Despite this model-independent construction, there is a precise mapping that can be worked out between the EFT action and the action of  any given single field DE/MG model that introduces a single scalar field and has a well defined Jordan frame~\cite{Gleyzes:2013ooa,Bloomfield:2013efa}. Therefore, as we will elaborate in this paper,  the EFT formalism can be used in two ways: as a general model-independent framework to test the theory of gravity on large scales, studying the effects of the different operators and the constraints that can be put by data on their coefficients; as a unifying language to analyze specific single scalar field DE/MG models, once the chosen model is mapped into the EFT framework. 
We refer the reader to~\cite{Gubitosi:2012hu,Gleyzes:2013ooa,Bloomfield:2013efa} for a detailed discussion of the assumptions and limitations of this framework, an illustration of the mapping and a complete inventory of the models that can be cast in the EFT language. Here we shall highlight that one of the assumptions on which the EFT is built is the validity of the weak equivalence principle, as discussed in~\cite{Gubitosi:2012hu,Bloomfield:2012ff}, which limits is range of applicability to models for which a Jordan frame, where all matter minimally couples to gravity, can be defined.  Despite of the inherent limitations, the EFT framework includes most of the viable approaches to the phenomenon of cosmic acceleration that will undergo scrutiny with upcoming cosmological surveys. We shall mention, among others, the Horndeski class which includes quintessence, k-essence, $f(R)$, covariant Galileon, the effective $4D$ limit of DGP~\cite{Nicolis:2008in} and more.

We modify the publicly available \emph{Code for Anisotropies in the Microwave Background} (CAMB)~\cite{CAMB,Lewis:1999bs}, creating what we will refer to as EFTCAMB, which will be publicly released soon. The latter is a full Einstein-Bolztmann code that can be used  to investigate the implications of the different  EFT operators on linear perturbations as well as  to study perturbations in any specific dark energy or modified gravity model that can be cast into the EFT language, once the mapping is worked out. Such a  code will be of great use for upcoming cosmological surveys, such as Euclid~\cite{euclid,Amendola:2012ys},  that aim at testing the underlying theory of gravity on large scales.

One of the virtue of the code is that it does not rely on any quasi-static (QS) approximation but still allows for the implementation of specific single field models of DE/MG. When fitting to data or performing forecasts for upcoming surveys, one generally focuses on sub-horizon scales and neglects the time derivatives of the gravitational potentials and scalar fields w.r.t. their spatial gradients, i.e. one assumes the QS regime. In Fourier space, this brings the Einstein and scalar field equations to  an algebraic form and simplifies significantly both the theoretical and the numerical setup. A widely used parametrization of modified gravity that relies on the QS approximation is the one introduced in~\cite{Bertschinger:2008zb} and commonly referred to as the BZ parametrization. While the QS description of the growth of structure generally gives a good representation of the evolution on sub-horizon scales (see e.g.~\cite{Hojjati:2012rf} for an analysis in $f(R)$ gravity), and significantly 
reduces the computing time, 
it might loose out on some dynamics at redshifts and scales that would leave an imprint within the reach of some ongoing and upcoming 
surveys~\cite{Lombriser:2013aj,Noller:2013wca}.  At the level of model-independent tests of gravity, implementations that do not employ 
the QS approximation are the Parametrized Post-Friedmann (PPF) modules of~\cite{Hu:2007pj,Fang:2008sn} as well as 
MGCAMB~\cite{Zhao:2008bn,Hojjati:2011ix}. The former uses a full set of equations for all linear scales, 
obtained by the interpolation between the super-horizon and the QS regime and it relies on three free functions and one parameter; 
however, in order to study specific models, one needs to work out interpolations and fits to the these functions and parameters for each case. 
The latter relies on a generic parametrization of the Poisson and anisotropy equation to form a complete 
and general set of equations for all linear scales, allowing for model-independent analysis of modified 
growth such as those of~\cite{Hojjati:2011xd,Zhao:2010dz,Asaba:2013xql}; however, one has to restrict to the QS regime in order to study a specific model. 
We shall mention also ISiTGR~\cite{Dossett:2011tn,Dossett:2012kd}, which is an integrated set of modified modules for use in testing whether observational data are consistent with general relativity on cosmological scales.
EFTCAMB is a full Einstein-Boltzmann code which does not rely on any QS approximation and is very general in terms of models and parametrizations that it can handle. We will illustrate the importance of allowing for full dynamics in view of upcoming data in Sec.~\ref{designer_fR_2}, by showing how, for some models, the time dependence of the scalar d.o.f. can have a non-negligible effect on the lensing of the Cosmic Microwave Background (CMB) detected by \emph{Planck}~\cite{Ade:2013tyw}.

Our code solves the full Klein-Gordon equation for the St$\ddot{\text{u}}$ckelberg field, which in the EFT formalism  encodes the departures from the standard cosmological model, as opposed to macroscopic hydrodynamic/fluid treatments~\cite{Sawicki:2012re,Bloomfield:2013cyf,Battye:2013ida}. This allows us to maintain an approach which is closer to the true nature of the theory as well as to have a direct control and easier interpretation of the possible instabilities related to this d.o.f. as we discuss in Sec.~\ref{Instabilities}. Furthermore, with our method we can easily evolve perturbations in models that cross the phantom-divide, as we will illustrate in Sec.~\ref{pure_EFT}. 

We focus on cosmological observables of interest for ongoing and upcoming surveys, 
in particular showing outputs of our code for the CMB temperature-temperature auto-correlation, 
the CMB lensing potential auto-correlation, the cross-correlation between temperature and lensing potential for the CMB and the matter power spectrum. 
We consider several models. To start with, we focus on $f(R)$ models that reproduce a $\Lambda$CDM expansion history and we compare our outputs to those of the common implementation of these theories in 
MGCAMB~\cite{Zhao:2008bn,Hojjati:2011ix,Hojjati:2012rf}. This allows us  to perform a consistency check of our code as well as to identify some peculiar features in the power spectra contributed by the sub-horizon dynamics of the scalaron, which is neglected by the QS implementation of $f(R)$ in MGCAMB. 
We then extend to designer $f(R)$ models with more general expansion histories, considering both a constant but different than $-1$ and a time-varying dark energy equation of state. We analyze in details all the imprints of these models on the different observables. Finally, we switch gears and instead of adopting a known model of modified gravity, we study some ghost-free power law parameterizations of the EFT background functions that display a phantom-divide crossing background.
The examples that we present should highlight the versatility of our EFTCAMB code: it can be used to 
evolve the full dynamics of linear perturbations for any given DE/MG model that can be cast into the EFT language, without the need to resort to the QS approximation; it provides a powerful and versatile tool to implement the EFT formalism as a model-independent parametrization to test gravity with large scale structure;  it allows us to investigate some poorly understood models which are permitted by the symmetries and stability conditions of EFT, such as the phantom-divide crossing ones. 

Finally, let us note that while our setup takes into account all contributions from operators that are at most quadratic in the perturbations,  for the numerical analysis of this paper we focus on models that involve only background operators, 
leaving the analysis with second order terms for future work. For a previous investigation of cosmological implications on a subset of models within the framework of effective field theory of cosmic acceleration see~\cite{Mueller:2012kb}. 
  
The paper is organized as follows. In Sec.~\ref{Sec:EFT} we review the effective field theory of dark energy framework and  we discuss both its formulation in unitary gauge and the one in which the scalar field is manifest  via the St$\ddot{\text{u}}$ckelberg trick. In Sec.~\ref{Sec:background} we describe how EFTCAMB deals with the background cosmology. In Sec.~\ref{Sec:lin_perturbations}, we present the equations for scalar linear perturbations, we discuss some general theoretical requirements for the stability of the theory and review the cosmological observables of interest. In Sec.~\ref{Sec:numerical} we present numerical results, including an in depth comparison of our outputs to those of MGCAMB for $f(R)$ models, new results for designer $f(R)$ with time-varying dark energy equation of state and some EFT parametrizations with a phantom-divide crossing. We conclude in Sec.~\ref{Sec:conclusion}.

 \section{Effective Field Theory of Dark Energy}\label{Sec:EFT}
 We define the action of the effective field theory of dark energy~\cite{Gubitosi:2012hu,Bloomfield:2012ff} in unitary gauge, Jordan frame and conformal time, as follows: 
\begin{align}\label{action}
  S = \int d^4x &\sqrt{-g}   \left \{ \frac{m_0^2}{2} \l[1+\Omega(\tau)\r]R+ \Lambda(\tau) - a^2c(\tau) \delta g^{00}+\frac{M_2^4 (\tau)}{2} (a^2\delta g^{00})^2 - \frac{\bar{M}_1^3 (\tau)}{2}a^2 \delta g^{00} \delta {K}\indices{^\mu_\mu} - \frac{\bar{M}_2^2 (\tau)}{2} (\delta K\indices{^\mu_\mu})^2 \right.
\nonumber \\ 
& \left. - \frac{\bar{M}_3^2 (\tau)}{2} \delta K\indices{^\mu_\nu} \delta K\indices{^\nu_\mu}
+\f{a^2\hat{M}^2(\tau)}{2}\delta g^{00}\delta R^{(3)} + m_2^2 (\tau) (g^{\mu \nu} + n^\mu n^\nu) \partial_\mu (a^2g^{00}) \partial_\nu(a^2 g^{00})
+ \ldots  \right\} + S_{m} [\chi_i ,g_{\mu \nu}],\nonumber\\
\end{align}
where $m_0$ is the bare Planck mass, $R$ is the Ricci scalar, $a^2\delta g^{00}=a^2g^{00}+1$,  $\delta R^{(3)}$, $\delta K_{\mu\nu}$ and $\delta K\indices{^\mu_\mu}$ are the perturbations respectively  to the upper time-time component of the metric, the three dimensional spatial Ricci scalar, the extrinsic curvature and its trace. The functions $\Omega$, $\Lambda$ and $c$ are free functions of the conformal time coordinate $\tau$ and are the only ones which affect the background dynamics, hence the name background operators. The vector $n^{\mu}$ is the normal to surfaces of constant time. $M_2$, $\bar{M}_1$, $\bar{M}_2$, $\bar{M}_3$, $\hat{M}_2$ and $m_2$  are functions of time with dimensions of  mass. These are second order operators which contribute only to the equations for perturbations. The ellipses in the action stand for higher order terms. Our notation follows more closely that one of~\cite{Bloomfield:2012ff}, however we work in conformal time and we multiply the Ricci scalar by $1+\Omega$ instead of simply $\Omega$ for reasons of accuracy in the numerical calculations that we will perform. $S_{m}$ is the action for all matter fields. We work in the Jordan frame with $1+\Omega$ parametrizing the conformal coupling to gravity, while in the Einstein frame the same function would describe the coupling of the scalar d.o.f. to matter. The EFT formalism is based on the assumption of the weak equivalence principle which ensures the existence of a metric universally coupled to matter fields and therefore of a well defined Jordan frame~\cite{Gubitosi:2012hu,Bloomfield:2012ff}.

For a detailed explanation of how this action~(\ref{action}) is constructed we refer the reader to~\cite{Creminelli:2008wc,Gubitosi:2012hu}. Here we shall briefly mention that it is built out of the only terms that are consistent with the unbroken symmetries of the theory, i.e. time-dependent spatial diffeomorphisms, and are organized in powers of the number of perturbations and derivatives. Such action encompasses all single scalar field dark energy and modified gravity models, minimally and non-minimally coupled. A matching between EFT  functions and specific single scalar field DE/MG models has been provided in~\cite{Gubitosi:2012hu,Bloomfield:2012ff,Bloomfield:2013efa,Gleyzes:2013ooa}.  

\subsection{The St$\ddot{\text{u}}$ckelberg field}
In action~(\ref{action}) the extra dynamical scalar d.o.f. that one expects both in modified gravity and dark energy models, is hidden inside the metric. The unitary gauge is particularly suited for the construction of the most general action to describe all single field dark energy and modified gravity model. However, in order to study the dynamics of perturbations it is more practical to disentangle the scalar d.o.f from the metric ones. This can be achieved via the St$\ddot{\text{u}}$ckelberg technique. Operationally, one restores the time diffeomorphism invariance by mean of an infinitesimal time coordinate transformation which introduces a scalar field, commonly referred to as the St$\ddot{\text{u}}$ckelberg field, that realizes the symmetry nonlinearly. We work in conformal time, so in our case the time diffeomorphism reads:
\begin{equation}
 \tau \rightarrow \tau + \pi(x^{\mu}),
\end{equation}
while spatial coordinates are left unchanged. In the context of EFT of Inflation~\cite{Cheung:2007st}, the St$\ddot{\text{u}}$ckelberg field is associated to the Goldstone boson, while in the effective field theory of dark energy the time translation invariance is no longer realized by the Goldstone scalar mode. Under this procedure,  time-dependent functions are modified according to 
\begin{equation}\label{taylorexpansionstuck}
f(\tau)\rightarrow f(\tau+\pi(x^{\mu}))=f(\tau)+\dot{f}(\tau)\pi+\frac{\ddot{f}(\tau)}{2}\pi^2+\dots
\end{equation}
and are typically Taylor expanded in $\pi$. Throughout the paper dots indicate derivations w.r.t. conformal time. Furthermore, operators that are not fully diffeomorphism invariant transform according to the tensor transformation law, generating dynamical terms for $\pi$. For a complete description see~\cite{Cheung:2007st,Gubitosi:2012hu,Bloomfield:2012ff}. To keep the resulting action at second order in perturbations, the functions multiplying the background operators,  i.e. $\{\Omega,\Lambda,c\}$, are Taylor expanded up to the second order, while the functions multiplying second (and higher) order operators are expanded to zero order.

Let us illustrate how the mechanism works for the background operators, by giving the explicit form of the resulting action in terms of the St$\ddot{\text{u}}$ckelberg field, up to second order in perturbations
\begin{align}\label{action_Stuck}
 S = \int d^4x \sqrt{-g}& \left \{ \frac{m_0^2}{2} \l[1+\Omega(\tau+\pi)\r]R+ \Lambda(\tau+\pi) - c(\tau+\pi)a^2\left[ \delta g^{00}-2\frac{\dot{\pi}}{a^2} + 2\hub\pi\left(\delta g^{00}-\frac{1}{a^2}-2\frac{\dot{\pi}}{a^2}\right) +2\dot{\pi}\delta g^{00} \right.\right. \nonumber \\
 &\left.\left.+2g^{0i}\partial_i\pi-\frac{\dot{\pi}^2}{a^2}+ g^{ij}\partial_i \pi \partial_j\pi -\l(2\hub^2+\dot{\hub}\r)\frac{\pi^2}{a^2} \right]+\cdots \right\}+ S_{m} [g_{\mu \nu}],
\end{align}
 where the scale factor has already been Taylor expanded according to Eq.~(\ref{taylorexpansionstuck}). We give the explicit expression of the contributions to action~(\ref{action_Stuck}) from second order operators in Appendix~\ref{higher_orders}. We will use action~(\ref{action_Stuck}) to derive the linearly perturbed Einstein equations in Sec.~\ref{Sec:lin_perturbations}. Please notice that our St$\ddot{\text{u}}$ckelberg field is defined w.r.t. to conformal time, therefore there is a factor of $a$ of difference w.r.t. the St$\ddot{\text{u}}$ckelberg field of~\cite{Gubitosi:2012hu,Bloomfield:2012ff}.
 
\section{The background}\label{Sec:background}

Varying the background part of the action~(\ref{action}) or, equivalently~(\ref{action_Stuck}), with respect to the metric and assuming a flat FLRW metric one obtains the following equations:
\ba\label{Fr1}
&&\hub^2=\f{a^2}{3m_0^2(1+\Omega)}(\rho_m+2c-\Lambda)-\hub\f{\dot{\Omega}}{1+\Omega},\\
\label{acc1}
&&\dot{\hub}=-\f{a^2}{6m_0^2(1+\Omega)}\l(\rho_m+3P_m\r)-\f{a^2(c+\Lambda)}{3m_0^2(1+\Omega)}-\frac{\ddot{\Omega}}{2(1+\Omega)},
\ea
where $\hub=\dot{a}/a$ and $\rho_m$ and $P_m$ are, respectively, the energy density and pressure of the matter components, for which we assume a perfect fluid form with the standard continuity equation:
\be\label{cont_matter}
\dot{\rho}_m=-3\hub(\rho_m+P_m). 
\ee 
Eqs.~(\ref{Fr1})-(\ref{acc1}) can be recast in the following, more conventional, form
\ba
\label{Friedmann}
&&\hub^2=\f{a^2}{3m_0^2(1+\Omega)}\l(\rho_m+\rho_Q\r),\\
\label{acceleration}
&&\dot{\hub}=-\f{a^2}{6m_0^2(1+\Omega)}\l(\rho_m+3P_m+\rho_Q+3P_Q\r),
\ea
if one defines
\ba\label{Q_quantities}
&&\rho_Q\equiv2c-\Lambda-\f{3m_0^2\hub\dot{\Omega}}{a^2},\nonumber\\
&&P_Q\equiv\Lambda+\f{m_0^2}{a^2}\l(\ddot{\Omega}+\hub\dot{\Omega}\r).
\ea
The latter can be interpreted as, respectively, the energy density and pressure of the effective dark fluid. Combining Eqs.~(\ref{cont_matter}),~(\ref{Friedmann}) and Eq.~(\ref{acceleration}) one obtains the following continuity equation for this dark component:
\be\label{conservationQ}
\dot{\rho}_Q=-3\hub\l(\rho_Q+P_Q\r)+\f{3m_0^2}{a^2}\hub^2\dot{\Omega},
\ee
which tells us that the energy density $\rho_Q$ is conserved only in the case of $\Omega={\rm const.}$, i.e. a minimally coupled theory.

The EFT formalism offers a bridge between theory and data which can be used to test gravity on large scales via the dynamics of linear cosmological perturbations. To this extent, one can use a designer approach to fix a priori the background evolution and use Eqs.~(\ref{Fr1})-(\ref{acc1}) to determine two of the EFT background functions $\{\Omega,\Lambda,c\}$ in terms of the third one. It turns out to be convenient to solve for $c$ and $\Lambda$ in terms of $\Omega$:
\ba
\label{cdesigner}
&&c=-\f{m_0^2\ddot{\Omega}}{2a^2}+\f{m_0^2\hub\dot{\Omega}}{a^2}+ \frac{m_0^2(1+\Omega)}{a^2}(\hub^2-\dot{\hub})-\frac{1}{2}(\rho_m+P_m),\\
\label{Lambdadesigner}
&&\Lambda= -\f{m_0^2\ddot{\Omega}}{a^2}-\f{m_0^2\hub\dot{\Omega}}{a^2}-\frac{m_0^2(1+\Omega)}{a^2}(\hub^2+2\dot{\hub})-P_m.
\ea

\subsection{Code implementation of the background cosmology}\label{Sec:designer}

As stated in the Introduction, we envisage our code to serve two purposes. One being the application of the EFT framework in a model-independent way, to study the effect of the different operators in action~(\ref{action}) on the dynamics of linear perturbations; and eventually constrain the time-dependent coefficients multiplying these operators. And the second one, having a versatile full Boltzmann code to study the evolution of perturbations in virtually any single field dark energy and modified gravity model for which a mapping to the EFT formalism can be worked out. The twofold nature of the code translates into the following two different procedures for the implementation of the background:
\begin{itemize}[leftmargin=*]
\item \emph{pure} EFT:  we fix the expansion history and we choose a viable form for $\Omega(\tau)$~\cite{Frusciante:2013zop}. We then use the EFT designer approach discussed above to get $c,\Lambda$ and either~(\ref{Q_quantities}) or the prescription described below (see Eq.~(\ref{rho_P_Q})) to get the Q quantities.  

\item  \emph{mapping} EFT: we focus on a particular dark energy or modified gravity model, e.g. $f(R)$; in this case the optimal way to proceed is to solve the background equations of the chosen model (which could involve a designer approach) and then use the mapping to the EFT formalism as described  in~\cite{Gubitosi:2012hu,Bloomfield:2012ff,Gleyzes:2013ooa,Bloomfield:2013efa} to reconstruct the corresponding EFT  functions and Eqs.~(\ref{Q_quantities}) to obtain the Q quantities~(\ref{Q_quantities}).
\end{itemize}
We will give explicit examples of the two cases above in Sec.~\ref{Sec:numerical}, when we present numerical results of our code.

For the actual implementation of the \emph{pure} EFT cases, we fix the expansion history to 
\be\label{Friedmann_designer}
 \mathcal{H}^2 = \frac{8\pi G}{3} a^2 (\rho_m + \rho_{\rm DE}),
 \ee
with
\be\label{density_DE}
\rho_{\rm DE}= 3H_0^2M_P^2\Omega_{\rm DE}^0 \, \exp \l[-3 \int_{1}^{a}\l(1+w_{\rm DE}(a')\r) \,d\ln{a}' \r],
 \ee
where $w_{\rm DE}$ represents the equation of state of the effective dark energy component and can be set accordingly to the model that one wants to study. In particular we will consider the following three cases:
\begin{itemize}
\item[] - $w_{\rm DE}=-1$, corresponding to a $\Lambda$CDM expansion history;
\item[] -  $w_{\rm DE} = \text{const} \neq -1$, we will refer to this case as  $w$CDM;
\item[] -  $w_{\rm DE}(a)=w_0 + w_a(1-a)$, i.e. the CPL parametrization~\cite{Chevallier:2000qy,Linder:2002et}, where $w_0$ and $w_a$ are constant, respectively the value and the derivative of $w_{\rm DE}$ today. 
\end{itemize}
From a comparison of~(\ref{Friedmann}),~(\ref{acceleration}) with~(\ref{Friedmann_designer}),~(\ref{density_DE}), one obtains the following correspondence:
\ba
\label{rho_P_Q}
&&\rho_Q=\l(1+\Omega\r)\rho_{\rm DE}+\Omega\rho_m\, , \nonumber\\
&& P_Q=\l(1+\Omega\r)P_{\rm DE}+\Omega P_m\,.
\ea
After fixing $w_{\rm DE}$, we use~(\ref{rho_P_Q}) to determine the Q quantities; we then choose an  $\Omega(\tau)$ and use~(\ref{cdesigner}) and~(\ref{Lambdadesigner}) to get $c$ and $\Lambda$. Let us note that the quantity $\rho_{\rm DE}$ represents one possible way of modeling the contribution of the dark component, alternative to the quantity $\rho_Q$ introduced above. The $Q$ and $DE$ quantities coincide in the case $\Omega=0$, i.e. when the dark sector is minimally coupled to gravity. However, when $\Omega\neq 0$, Eq.~(\ref{Q_quantities}) gives a more proper representation of the effective scalar d.o.f.  of the dark sector, taking into account the coupling to matter and the corresponding exchange of energy between the dark and the matter sectors. In fact, the continuity equation~(\ref{conservationQ}) and that one for~(\ref{density_DE}) coincide when $\dot{\Omega}= 0$, while for $\dot{\Omega}\neq 0$ the density $\rho_Q$ receives an extra contribution from the coupling to matter. 
We choose to formulate the designer approach for our code in terms of $(\rho_{\rm DE},w_{\rm DE})$, which allows for a more direct implementation of the background cosmology in CAMB. However, we express the equations for linear perturbations in terms of the $Q$ quantities, as it is usually done in the EFT framework, since those better represent the contributions to the 
evolution of perturbations from the EFT dark component.
 
  While in the linearly perturbed equations of next Section we keep $c$ and the Q quantities, we implicitly assume that once the background is fixed, those will be expressed  in terms  of the expansion history and $\Omega$ via a combination of~(\ref{cdesigner}),~(\ref{Lambdadesigner}) and~(\ref{rho_P_Q}) for the \emph{pure} EFT cases, and via the matching recipe and~(\ref{Q_quantities}) in the \emph{mapping} EFT cases.
 
Before concluding this Section, we shall comment on a fundamental difference between the \emph{pure} and \emph{mapping} EFT cases. In the former, the designer approach serves solely the purpose of fixing the background, and therefore only the EFT functions $\{\Omega,\Lambda,c\}$; when studying the dynamics of linear perturbations one needs to independently choose a form for the EFT functions multiplying second order operators, (which of course includes the case in which all those coefficients are set to zero). In the mapping case instead, once the model is chosen and the corresponding background is solved, one can reconstruct \emph{all} the time-dependent coefficients in the EFT action through the matching procedure, including the higher order ones if the model under consideration involve them. Therefore in the mapping case, once the model is specified one has all the necessary ingredients to study the dynamics of cosmological perturbations.
 
\section{Scalar linear perturbations}\label{Sec:lin_perturbations}
We shall now derive the linearly perturbed Einstein equations that are needed in order to evolve scalar perturbations in CAMB. We work in synchronous gauge with the line element given by
\be\label{metric}
ds^2 = a(\tau)^2\left[-d\tau^2 + \left(\delta_{ij} + h_{ij}dx^idx^j\right)\right],
\ee
where the scalar mode of $h_{ij}$ in Fourier space can be decomposed into 
\be
h_{ij} = \int{dk^3e^{i\textbf{k}\cdot\textbf{x}}\left[\hat{k}_i\hat{k}_j h(k,\tau) + \left(\hat{k}_i\hat{k}_j-2\delta_{ij}\eta(k,\tau) \right)\right]},
\ee
with $h$ denoting the trace of $h_{ij}$. Unless explicitly stated otherwise, we work with Fourier transforms of all cosmological perturbations.
 
While the functions $\{\Omega,\Lambda,c\}$ are the only ones affecting the background dynamics in the EFT formalism, when we move to linear perturbations, more operators come into play; indeed, all the remaining functions in action~(\ref{action}), or equivalently~(\ref{full_action_Stuck}), that multiply second order operators, will also affect the dynamics of linear perturbations. For the sake of brevity, here we focus on the terms contributed by background operators and we list the contributions from second order operators in Appendix~\ref{higher_orders}. 
 
Starting from the action in terms of the St$\ddot{\text{u}}$ckelberg field~(\ref{action_Stuck}), and simplifying the background terms, to linear order in scalar perturbations we have:
\\
\\
\underline{time-time} Einstein equation:
\ba\label{00}
k^2\eta=-\f{a^2}{2m_0^2(1+\Omega)}\l[\delta\rho_m+\dot{\rho}_Q\pi+2c\l(\dot{\pi}+\hub\pi\r)\r]+\l(\hub+\f{\dot{\Omega}}{2(1+\Omega)}\r)k\mathcal{Z}+\f{\dot{\Omega}}{2(1+\Omega)}\l[3(3\hub^2-\dot{\hub})\pi+3\hub\dot{\pi}+k^2\pi\r]\,,\nonumber\\
\ea
\underline{momentum} Einstein equation:
\be\label{0i}
\f{2}{3}k^2\l(\sigma_*-\cal{Z}\r)=\f{a^2}{m_0^2(1+\Omega)}\l[(\rho_m+P_m)v_m+(\rho_Q+P_Q)k\pi\r]+k\f{\dot{\Omega}}{(1+\Omega)}\l(\dot{\pi}+\hub\pi\r) \,,
\ee
\underline{space-space off-diagonal} Einstein equation:
\be\label{ijoff}
k\dot{\sigma}_*+2k\hub\sigma_*-k^2\eta=-\frac{a^2P\Pi_m}{m_0^2(1+\Omega)}-\frac{\dot{\Omega}}{(1+\Omega)}\l(k\sigma_*+k^2\pi\r),
\ee
\underline{space-space trace} Einstein equation:
\ba\label{space-space_trace}
\ddot{h}&=&-\f{3a^2}{m_0^2(1+\Omega)}\l[\delta P_m+\dot{P}_Q\pi+\l(\rho_Q+P_Q\r)\l(\dot{\pi}+\hub\pi\r)\r]-2\l(\f{\dot{\Omega}}{1+\Omega}+2\hub \r)k\mathcal{Z}+2k^2\eta\nonumber\\
&&-3\f{\dot{\Omega}}{(1+\Omega)}\l[\ddot{\pi}+\l(\f{\ddot{\Omega}}{\dot{\Omega}}+3\hub\r)\dot{\pi}+\l(\hub\f{\ddot{\Omega}}{\dot{\Omega}}+5\hub^2+\dot{\hub}+\f{2}{3}k^2\r)\pi\r]\,,
\ea

\underline{$\pi$} field equation:
\ba\label{pi0}
c\ddot{\pi}+(\dot{c}+4\hub c)\dot{\pi}+\l[\frac{3}{2}\frac{m_0^2\dot{\Omega}}{a^2}(\ddot{\hub}-2\hub^3)-2\dot{\hub}c+\hub\dot{c}+6\hub^2c+k^2c\r]\pi+ck\mathcal{Z}-\frac{m_0^2\dot{\Omega}}{4a^2}\l[\ddot{h}-4k^2\eta+6k\hub\mathcal{Z}\r]=0,
\ea
where $2k{\mathcal Z}\equiv \dot{h}$ and $2k\sigma_*\equiv \dot{h}+6\dot{\eta}$ are the standard CAMB variables~\cite{CAMBNotes}. As we will discuss shortly in Sec.~\ref{Instabilities}, it is important to demix the degrees of freedom in order to perform the appropriate stability analysis of perturbations in the dark sector~\cite{Gubitosi:2012hu}. Namely, one shall substitute for $\eta$ and $\ddot{h}$ using Eq.~(\ref{00}) and~(\ref{space-space_trace}), respectively, in order to obtain the following equation:

\ba\label{pi}
&&\l(c+\f{3m_0^2}{4a^2}\f{\dot{\Omega}^2}{(1+\Omega)}\r)\ddot{\pi}+\l[\f{3m_0^2}{4a^2}\f{\dot{\Omega}}{(1+\Omega)}\l(\ddot{\Omega}+4\hub\dot{\Omega}+\f{(\rho_Q+P_Q)a^2}{m_0^2}\r)+\dot{c}+4\hub c-\f{\dot{\Omega}}{2(1+\Omega)}c\r]\dot{\pi}\nonumber\\
&&+\l[\f{3}{4}\f{m_0^2}{a^2}\f{\dot{\Omega}}{(1+\Omega)}\l(\f{(3\dot{P}_Q-\dot{\rho}_Q+3\hub(\rho_Q+P_Q))a^2}{3m_0^2}+\hub\ddot{\Omega}+8\hub^2\dot{\Omega}+2(1+\Omega)(\ddot{\hub}-2\hub^3)\r)\right.\nonumber\\
&&\left.-2\dot{\hub}c+\l(\dot{c}-\f{\dot{\Omega}}{2(1+\Omega)}c\r)\hub+6\hub^2 c+\l(c+\f{3m_0^2}{4a^2}\f{\dot{\Omega}^2}{(1+\Omega)}\r)k^2\r]\pi\nonumber\\
&&+\l[c+\f{3}{4}\f{m_0^2}{a^2}\f{\dot{\Omega}^2}{(1+\Omega)}\r]k{\mathcal Z} +\f{1}{4}\f{\dot{\Omega}}{(1+\Omega)}\l(3\delta P_m-\delta\rho_m\r)=0.
\ea
In our numerical code, we set the standard initial conditions for matter components and curvature perturbations in the radiation dominated epoch,  at a time when the corresponding momentum mode re-enter the horizon. For the St$\ddot{\text{u}}$ckelberg field instead, we set initial conditions at a later time, corresponding to $a_{\pi}=0.01$.
The reasons for this choice are several. First of all, we are interested in the late time accelerating universe and we typically want our theory  to reproduce standard GR at early times ($a<a_{\pi}$). 
In other words, we expect  the St$\ddot{\text{u}}$ckelberg field not to be excited at early times. This fact also makes initial conditions for this scalar field  less motivated at deep redshift, when the other matter components initial conditions are  instead well defined.
Finally, from the numerical point of view, the system is more easily controlled since, not evolving the $\pi$ equation at early times, we avoid some potential high frequency dynamics that would make the integration time longer, while keeping track of the underlying mode of evolution of the scalar field. Indeed, since the equation of motion for the St$\ddot{\text{u}}$ckelberg field,~(\ref{pi}), is coupled to metric and matter perturbations, which behaves as an external driving source, we set the $\pi$ field to trace the dynamics of the source at times earlier than $a_{\pi}$. In this way we can get regular and proper initial conditions for the $\pi$ field at $a_{\pi}$, while avoiding potential high frequency dynamics around the underlying growing mode which anyhow are not expected to leave imprints on physical observables.

\subsection{Stability of perturbations in the dark sector}\label{Instabilities}
In this subsection we shall focus on some requirements for theoretical stability that can be enforced on the EFT functions to ensure that the underlying gravitational theory is acceptable. To  this purpose we implement in our code a consistency check for the fulfillment of such stability conditions. For the following discussion it is more convenient to write the $\pi$ field equation as follows:
\be\label{pi_inst}
A(\tau)\,\ddot{\pi} + B(\tau)\,\dot{\pi} + C(\tau)\,\pi + k^2\,D(\tau)\,\pi + E(\tau,k) =0
\ee
where the coefficients $\{A,\dots,E\}$ can be easily read from Eq.~(\ref{pi}) (and the results of Appendix~\ref{higher_orders} if second order operators are at play. 
In that case also $A$, $B$ and D may display $k-$dependence). 
Relying on the discussion of~\cite{Creminelli:2008wc}, we place the following theoretical constraints:
\begin{itemize}
\item $1+\Omega > 0$:  this condition on the non-minimal coupling function is required in order to ensure that the effective Newtonian constant does not change sign. Violating this condition, classically, would imply a Universe quickly becoming inhomogeneous and anisotropic~\cite{Nariai:1973eg,Gurovich:1979xg}, while at the quantum level it will correspond to the graviton turning into a ghost~\cite{Nunez:2004ji};
\item $A>0$: this second condition follows from requiring that our effective scalar d.o.f. should not be a ghost, i.e. the corresponding kinetic energy term should be positive. At the classical level there is no serious danger in this situation while at the quantum level the underlying physical theory can show instability of the vacuum~\cite{Cline:2003gs};
\item $c_s^2 \equiv D/A \leq 1$: the third condition ensures that the sound speed of $\pi$ does not exceed the speed of light to prevent scalar perturbations from propagating super-luminary. This condition is no longer true when treating, for example, Lorentz violating theories~\cite{Liberati:2013xla};
\item $m_\pi^2 \equiv C/A \geq 0$: last, we enforce the mass of the scalar d.o.f. to be real~\cite{Gumrukcuoglu:2013nza}, to avoid tachyonic instabilities. In $f(R)$ gravity, that we consider in Sec.~\ref{fR_numerical}, this condition is necessary to guarantee a stable high-curvature regime~\cite{Pogosian:2007sw}.
\end{itemize}
The above conditions could be relaxed in  certain cases depending on the specific theory of gravity one is interested in and, of course, our code can be easily edited to check different stability requirements. Let us briefly comment on this. The first two conditions are quite general and can be relaxed just in elaborated models that can associate a physical meaning to the negative branch of $A$ and $1+\Omega$.  Furthermore, their positive and negative branches are disconnected so that no theory can allow these two quantities to cross zero as this will violate the mathematical consistency of the initial value problem for Eqs.~(\ref{Friedmann}) and~(\ref{pi}). The last two conditions are milder and more strictly related to the particular theory one wants to test.  Therefore, they can be relaxed in many ways if the EFT formalism  is used to test some peculiar model that naturally permits their violation. In this regards, we recall, among others, cosmological models which allow for viable DE models in Lorentz 
violating theories~\cite{Blas:2011en,
Audren:2013dwa} and rolling tachyon condensates~\cite{Padmanabhan:2002cp,Bagla:2002yn,Abramo:2003cp,Aguirregabiria:2004xd,Copeland:2006wr}. 
 
As pointed out in~\cite{Creminelli:2008wc}, there are other types of instabilities that can be studied efficiently within the EFT framework. We leave their thorough investigation for future work.

\subsection{Observables}
In view of using our code to test gravity with upcoming and future cosmological surveys, the observables of interest are all the two-point auto- and cross-correlations between Weak Lensing (WL), Galaxy Clustering (GC) and Cosmic Microwave Background (CMB) temperature and polarization anisotropy. We refer the reader to~\cite{Zhao:2008bn} for a thorough discussion of these observables and the corresponding angular power spectra. In this paper we show outputs of our code for the temperature-temperature auto-correlation, the CMB lensing potential auto-correlation, the cross-correlation between temperature and 
lensing potential for the CMB and the matter power spectrum. 

It is expected that the dynamics of the St$\ddot{\text{u}}$ckelberg field will mainly affect the  time evolution of the metric potentials and matter perturbations at late times. Therefore we expect to see the more noticeable effects in observables such as the Integrated Sachs-Wolfe (ISW) effect of the CMB and WL. The former is a secondary anisotropy induced by the time evolution of the Weyl potential 
($\psi\equiv(\Phi+\Psi)/2$ in Newtonian gauge \footnote{where we assume the following convention for the Newtonian gauge:~${\rm d}s^2=a^2(\tau)[-(1+2\Psi){\rm d}\tau^2 + (1-2\Phi){\rm d}x^2]$. The gauge transformations between Newtonian and synchronous gauges are given by: 
$\Psi = \dot\sigma_{\ast}/k+\mathcal H\sigma_{\ast}/k$,~$\Phi=\eta-\mathcal H\sigma_{\ast}/k$.}) at late times. The latter involves the distortion of light rays
when they pass close to clustering objects, such as galaxies and clusters; it is sourced by the spatial gradients of the Weyl potential.
During the accelerated epoch, no significant polarization modes of the CMB photon are generated, therefore we will not consider them here.

The CMB temperature angular spectrum can be computed via the line of sight integration method~\cite{Seljak:1996is}
\be
C_\ell^{TT}=(4\pi)^2\int \frac{{\rm d}k}{k}~\mathcal P(k)\Big|\Delta_\ell^{\rm T}(k)\Big|^2,\\
\ee
where $\mathcal P(k)=\Delta_{\mathcal R}^2(k)$ is the primordial power spectrum and the radiation transfer function
\be
\Delta_\ell^T(k)=\int_0^{\tau_0}{\rm d}\tau~e^{ik\mu(\tau-\tau_0)}S_{\rm T}(k,\tau)j_\ell[k(\tau_0-\tau)]
\ee
is sourced by
\begin{align}\label{T_source}
S_{\rm T}(k,\tau) =& e^{-\kappa}\left(\dot{\eta} + \frac{\ddot{\sigma_{\ast}}}{k} \right) 
+ g\left(\Delta_{\rm T,0} +2\frac{\dot{\sigma_{\ast}}}{k} +\frac{\dot{v}_{\rm B}}{k} + \frac{\Pi}{4} +\frac{3\ddot{\Pi}}{4k^2} \right) 
+ \dot{g}\left(\frac{\sigma_{\ast}}{k} + \frac{v_B}{k} +\frac{3\dot{\Pi}}{4k^2} \right) +\frac{3}{4k^2}\ddot{g}\Pi\;,
\end{align}
where $\tau_0$, $\mu$, $\kappa$, $g$, $\Delta_{\rm T,0}$, $v_{\rm B}$ and $\Pi$ are, respectively, the present conformal time, 
angular separation, optical depth, 
visibility function, intrinsic CMB density perturbations at the last scattering surface, 
velocity of baryonic matter and total anisotropic stress of normal matter (which includes CMB photons, 
massless/massive neutrino). Since the recombination of electrons and protons 
happens very fast, the visibility function $g$ peaks sharply at that early moment,
so we do not expect the St$\ddot{\text{u}}$ckelberg field to affect the terms proportional to the visibility function and its derivatives. 
As already discussed, the only relevant term of~(\ref{T_source}) for our analysis is the ISW one, which can be expressed as follows:
\begin{align}
\ddot{\sigma}_{\ast} + k\dot{\eta} =& -2 \mathcal{H}\dot{\sigma_{\ast}} 
-2 \dot{\mathcal{H}}\sigma_{\ast} +\frac{v_m}{1+\Omega}\frac{a^2(\rho_m+P_m)}{m_0^2} 
-\frac{1}{k(1+\Omega)} \frac{d}{d\tau}\left(\frac{a^2P}{m_0^2}\Pi\right) \nonumber \\
	& +\frac{k\pi}{1+\Omega}\frac{a^2 (\rho_Q + P_Q)}{m_0^2} 
	+ \frac{\dot{\Omega}}{1+\Omega}\bigg[k \mathcal{H}\pi 
	-\dot{\sigma}_{\ast} + \frac{1}{k(1+\Omega)}\frac{a^2 P}{m_0^2} \Pi \bigg]. \\
\end{align}
As for WL, we calculate its angular power spectrum following the convention of~\cite{CAMBNotes,Lewis:2006fu}:
\begin{align}
C_\ell^{\psi\psi}=4\pi\int \frac{{\rm d}k}{k}\mathcal{P}(k)\left[\int_0^{\chi_{\ast}}{\rm d}\chi~ S_{\psi}(k;\tau_0-\chi)j_\ell(k\chi)\right]^2,
\end{align}
where the source $S_{\psi}$ is given in terms of the transfer function of the Weyl potential $\psi$, i.e.:
\ba
&&S_{\psi}(k;\tau_0-\chi)=2T_{\psi}(k;\tau_0-\chi)\left(\frac{\chi_{\ast}-\chi}{\chi_{\ast}\chi}\right),\\
&&T_{\psi}(k,\tau) = \frac{\dot{\sigma_{\ast}} + k\eta}{2} =
\frac{1}{2}\left[-2\mathcal{H}\sigma_{\ast}
+2 k \eta -\frac{1}{k (1+ \Omega)}\frac{a^2 P}{m_0^2} \Pi
- \frac{\dot{\Omega}}{1+\Omega} \left(\sigma_{\ast} +
k\pi \right)\right]\;.
\ea
Conventionally, the line of sight integral in the lensing source is expressed in terms of comoving distance $\chi$.
Here $\chi_{\ast}$ is the comoving distance of the source objects. In this paper we will focus on CMB lensing, for which the source object is a single distant plane (since the electron-proton recombination is 
approximately instantaneous), i.e. $\chi_{\ast}$ corresponds to the comoving distance to last scattering surface.
At leading order, the relationship between  
comoving distance and conformal time reads $\chi=\tau_0-\tau$. 
Since  ISW and WL are sourced by the same potential, 
one being sensitive to time derivatives and the other to spatial gradients of the Weyl potential, 
it is expected that the two effects are strongly correlated and this correlation produces 
a non-zero cross-spectrum $C_\ell^{\rm T\psi}$~\cite{Hu:2000ee}:
\be
C_\ell^{\rm T\psi}=4\pi\int~\frac{{\rm d}k}{k}\mathcal P(k)\left\{\int_0^{\tau_0}~
{\rm d}\tau ~e^{ik\mu(\tau-\tau_0)}e^{-\kappa}(\dot\Phi+\dot\Psi)j_\ell\Big[k(\tau_0-\tau)\Big]\times
\int_{\tau_{\ast}}^{\tau_0}~{\rm d}\tau S_{\psi}(k;\tau)j_\ell\Big[k(\tau_0-\tau)\Big]\right\}\;,
\ee
with $\tau_{\ast}$ denoting for the conformal time at recombination. 

Finally, the matter power spectrum  can be computed via
\be
P(k)=\frac{2\pi^2}{k^3}\mathcal P(k)\Delta_{\rm T}(k)^2,
\ee
with the matter transfer function defined as~\cite{Eisenstein:1997ik}
\be
\Delta_{\rm T}(k)=\frac{\delta_m(k,z=0)\delta_m(0,z=\infty)}{\delta_m(k,z=\infty)\delta_m(0,z=0)}\;,
\ee
which describes the evolution of matter density perturbations through the epochs of horizon 
crossing and radiation/matter transition. A proper calculation of $\Delta_{\rm T}(k)$ requires that in our code we take all types
of non-relativistic matter into account, and follow the growth of each mode outside and inside the horizon.

\section{Numerical Results}\label{Sec:numerical}
In this Section we showcase the reliability and scope of EFTCAMB by comparing it with an existing code, as well as producing some interesting new results. While we have all the necessary ingredients to consider models which involve also second order operators in action~(\ref{action}), for the numerical analysis of this paper we focus on the cases that involve only the background operators. The examples that we present should convey the wide range of applicability of our code.

We will first focus on $f(R)$ models and  compare our code to the common implementation of these theories in MGCAMB~\cite{Zhao:2008bn,Hojjati:2011ix,Hojjati:2012rf}, restricting to $\Lambda$CDM expansion history. 
Then, we extend to designer $f(R)$ models with generic constant and time-varying dark energy equation of state in the part of this Section devoted to new results. 
As we will illustrate with an example in the $f(R)$ case, given the accuracy of ongoing and upcoming surveys, as well as the range of scales that they cover, in order to extract predictions about observables such as CMB lensing, it is important to employ a code that evolves the full dynamics of the system on linear scales, without employing the QS approximation. This is one of the qualities of our code which allows for an implementation of the full dynamics of a given model, without the need of reducing to the QS regime. 

EFTCAMB of course can be used also to fulfill the true purpose the EFT formalism has been envisaged for, i.e. a framework for model-independent tests of gravity on cosmological scales. To this extent, presumably one fixes the expansion history as discussed in~\cite{Gubitosi:2012hu,Bloomfield:2012ff} and briefly reviewed in Sec.~\ref{Sec:background}, and then focuses on the dynamics of cosmological perturbations studying the effects of the different operators in action~(\ref{action}). In this case it is necessary to select some parametrization for the functions of time multiplying the operators under consideration. 
Restricting to the background operators, we show the outputs for power law parameterization of the remaining EFT free function $\Omega(a)$
with a phantom-divide crossing background.

Throughout this paper we will always use the following cosmological parameters: $H_0=70\,\mbox{Km}/\mbox{s}/\mbox{Mpc}$, $\Omega_b = 0.05$, $\Omega_c=0.22$, $T_{\rm CMB}=2.7255\,\mbox{K}$. 

\subsection{$f(R)$ gravity: comparison and new results}\label{fR_numerical}
As an illustration of how EFTCAMB can be used in its \emph{mapping} EFT version, we shall perform a thorough analysis of $f(R)$ models. Let us start briefly reviewing the theory of $f(R)$ gravity. We consider the following Lagrangian in Jordan frame
\be\label{action_fR}
S=\int d^4x \sqrt{-g} \l[R+f(R)\r]+S_m \,,
\ee
where $f(R)$ is a generic function of the Ricci scalar and the matter sector is minimally coupled to gravity. These models can be mapped into the EFT formalism via the following matching~\cite{Gubitosi:2012hu}:
\begin{align}\label{fR_matching}
\Lambda=\frac{m_0^2}{2}\l[f- Rf_R \r] \hspace{0.5cm};\hspace{0.5cm} c=0 \hspace{0.5cm};\hspace{0.5cm} \Omega=f_R \,.
\end{align}

For a detailed discussion of the cosmology in $f(R)$ theories we refer the reader 
to~\cite{Song:2006ej,Bean:2006up,Pogosian:2007sw,DeFelice:2010aj}. Here we will briefly review the main features that are of interest for our analysis. The higher order nature of the theory translates into having an extra scalar d.o.f. which can be identified with the field $f_R\equiv df/dR$, commonly dubbed the \emph{scalaron}~\cite{Starobinsky:2007hu}. Implementing the matching to EFT, we have that St$\ddot{\text{u}}$ckelberg field for $f(R)$ theories is given by $\pi=\delta R/R$~\cite{Gubitosi:2012hu}, which can be easily related to the perturbation of the scalaron, $\delta f_R$.

Viable $f(R)$ models need to satisfy certain conditions of stability and consistency with local tests of gravity~\cite{Pogosian:2007sw}, which can be inferred from the conditions in Sec.~\ref{Instabilities} once the matching~(\ref{fR_matching}) is implemented. 
Finally, given the higher order of the theory, it is possible to reproduce any given expansion history by an appropriate choice of the $f(R)$ function~\cite{Song:2006ej,Pogosian:2007sw}. In other words, $f(R)$ models can be treated with the so called \emph{designer} approach which consists in fixing the expansion history and then using the Friedmann equation as a second order differential equation for $f[R(a)]$. As we will recap shortly, generically one finds a family of viable models that reproduce this expansion; the latter are commonly labeled by the boundary condition at present time, $f_R^0$. Equivalently, they can be parametrized by the present day value of the function:
\begin{equation}
\label{ComptonWave}
B=\frac{f_{RR}}{1+f_R}\frac{\hub \dot{R}}{\dot{\hub}-\hub^2}\,.
\end{equation}
Let us recall that the heavier the scalaron the smaller $B_0$ and $|f_R^0|$.

\subsubsection{Comparison with MGCAMB in a  $\Lambda$CDM background}\label{BZ_fR}
We shall start comparing our results for $f(R)$ theories with those of the publicly available MGCAMB code~\cite{Zhao:2008bn,Hojjati:2011ix}. Since we construct our code on the CAMB version of March 2013, in order to make a senseful comparison we use an updated version of MGCAMB based on the same version of CAMB and developed by the authors of~\cite{Hu:2013aqa}.   

MGCAMB relies on two functions of time and scale to parametrize deviations in the Poisson and anisotropy equations, closing  the system of equations for matter in conformal Newtonian gauge with the two following equations:
\be\label{mugamma}
k^2\Psi\equiv-\f{a^2}{2M_P^2}\mu(a,k)\rho_m\Delta_m,\hspace{2cm}\frac{\Phi}{\Psi}\equiv\gamma(a,k)\,.
\ee
In order to evolve perturbations in $f(R)$ models one has to specify the corresponding forms for $\mu(a,k)$ and $\gamma(a,k)$, and this can be achieved by taking the QS limit of the linearly perturbed equations, which corresponds to neglecting time-derivatives of the metric potentials and of the scalar field, as well as focusing on sub-horizon scales $k\gg \mathcal{H}$. In this limit we have:
\be\label{fR_QSA}
k^2\Psi=-\frac{1}{1+f_R}\frac{1+4f_{RR}/(1+f_R)k^2/a^2}{1+3f_{RR}/(1+f_R)k^2/a^2}\f{a^2\rho_m\Delta_m}{2M_P^2},\hspace{1.5cm}
\f{\Phi}{\Psi}=\frac{1+2f_{RR}/(1+f_R)k^2/a^2}{1+4f_{RR}/(1+f_R)k^2/a^2}.
\ee
On sub-horizon scales the dynamics of linear perturbations in $f(R)$ is generally described sufficiently well by this QS approximation~\cite{Hojjati:2012rf}. Eqs.~(\ref{fR_QSA}) have inspired the following parametrization~\cite{Bertschinger:2008zb,Giannantonio:2009gi,Hojjati:2012rf,Hojjati:2011ix}:
\be
\label{BZ}
\mu^{\rm BZ}(a,k)=\frac{1}{1-B_0 \Omega_m a^{s-1}/2}\frac{1+2/3B_0\l(k/H_0\r)^2a^s}{1+\f{1}{2}B_0\l(k/H_0\r)^2a^s},\hspace{1.5cm}
\gamma^{\rm BZ}(a,k)=\frac{1+1/3B_0\l(k/H_0\r)^2a^s}{1+2/3B_0\l(k/H_0\r)^2a^s},
\ee
to which we will refer as the \emph{BZ} parametrization that consists in assuming
\begin{align}
\frac{f_{\rm RR}}{1+f_{\rm R}}\equiv \frac{B_0}{6H_0^2}a^{s+2}.
\label{Eq:BZansatz}
\end{align}

A standard way of extracting predictions for cosmological observables and comparing $f(R)$ models to data is the one of modeling the late time universe by inserting Eq.~(\ref{BZ}) into MGCAMB, leaving $B_0$ as a free parameter and fixing $s=4$~\cite{Zhao:2008bn}. 
Let us recall that a $f(R)$ model defined by Eq.~(\ref{BZ}) with a constant value for $s$ will not in general be capable of reproducing the full $\Lambda$CDM expansion history. However, it works as a good approximation for each epoch alone~\cite{Thomas:2011pj}, as can be inferred from Eq.~(\ref{Eq:BZansatz}). Indeed a reasonable value of $s$ is given by $s\approx 5$ during radiation domination, $s\geq4$  during matter domination and $s<4$ during the late time phase of accelerated expansion.
For small values of $B_0$, it is customary to fix  $s=4$ as discussed in~\cite{Hojjati:2012rf}, however here we will re-examine this choice in view of the precision and extent of upcoming surveys.
\begin{figure}[t!]
\begin{center}
  \includegraphics[width=0.9\textwidth]{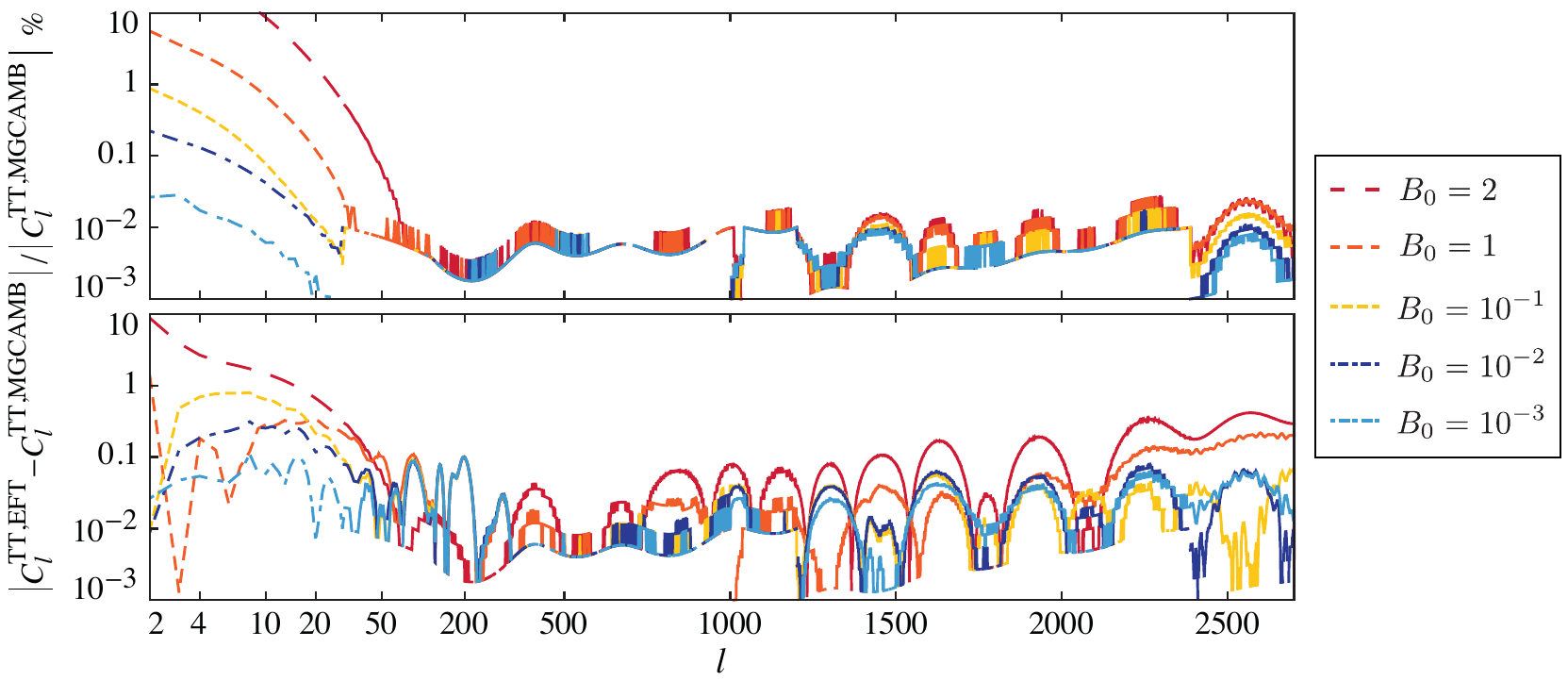}
  \caption{\label{Fig.Relative_Diff} \emph{Upper panel}: comparison between the temperature anisotropy angular power spectra of EFTCAMB and MGCAMB for $f(R)$ models with a $\Lambda$CDM expansion history but different values of $B_0$ and modeled via the BZ approach described in Sec.~\ref{BZ_fR}. \emph{Lower panel}: same comparison for the case of a designer $f(R)$ model with $\Lambda$CDM background. For a detailed interpretation of the plots see Sec.~\ref{BZ_fR}.} 
\end{center}
\end{figure}

In order to compute observables for these theories with MGCAMB it suffices to fix the expansion history to that of $\Lambda$CDM, $s=4$ and input~(\ref{BZ}) for $\mu$ and $\gamma$ for several choices of $B_0$. 
EFTCAMB, on the contrary, does not rely on the quasi-static BZ parametrization, but rather solves the full equations; therefore even after fixing the expansion history to $\Lambda$CDM we need to feed the code a form for the EFT  functions. We consider two cases:
\begin{itemize}[leftmargin=*]
\item BZ case: from Eq.~(\ref{Eq:BZansatz}) we read off the implications of the BZ parametrizations for $f(R)$ and then we reconstruct the corresponding $\Omega$ and $\Lambda$ to input in the full equations for linear perturbations. We again stress that our code does not rely on any QS approximation;
\item Designer case: we implement in our code the $f(R)$ designer approach to reconstruct viable $f(R)$ models that mimic the $\Lambda$CDM expansion history and determine the corresponding $\Omega,\Lambda$ via the matching, Eq.~(\ref{fR_matching}).
\end{itemize}
The BZ case allows us to make a check of reliability of our code with minimal changes with respect to the way $f(R)$ theories are treated in MGCAMB. The designer case corresponds to a proper full treatment of $f(R)$ models and therefore let us fully exploit the potential of our code, avoiding  spurious effects due to the BZ approximation; this will allow us to check the accuracy of the QS approximation in $f(R)$ models to a new extent. The latter case corresponds to the proper treatment of the background operators in the \emph{mapping} EFT cases.

Let us start with the BZ case. Using the matching formulae~(\ref{fR_matching}) we see that the BZ ans$\ddot{\text{a}}$tz~(\ref{Eq:BZansatz}) can be mapped into the EFT formalism as follows:
\begin{align}\label{BZEFTmatching}
\Omega &= -1 + e^{-\frac{3 B_0 \Omega_m a^{s-1}}{2(s-1)}}= -\frac{3}{2}\frac{B_0\Omega_m a^{s-1}}{s-1} +O(B_0^2), \nonumber\\
\frac{\Lambda}{m_0^2} &= -\frac{\rho_{\rm DE}}{m_0^2} + B_0H_0^2\frac{27a^4\Omega_m^2 - 9\Omega_m a^s \left(4 a^3 (s-4)\Omega_\Lambda +(s-1)\Omega_m \right)}{4 a^4 (s-4) (s-1)} +O(B_0^2).
\end{align}

As in MGCAMB, we fix $s=4$ and we use different values of $B_0$ ranging from very large ones ($B_0=2$) to very small ones ($B_0=10^{-3}$).
The comparison of the temperature spectra from the two codes is shown in the upper panel of Fig.~\ref{Fig.Relative_Diff}. 
As we can clearly see the agreement on small scales is very good ($\lesssim 0.01\%$) and remains under control ($\lesssim 0.1\%$) even on very large scales for small values of $B_0$ ($\lesssim 0.01$). 
We  get some tension between the two codes, (relative difference $> 1\%$), at low multipoles for large values of $B_0$ ($\gtrsim 0.1$). This is partially due to the way we treat the background in this case; first of all when $B_0\gtrsim 1$ the correction term in~(\ref{BZEFTmatching}) cannot be neglected anymore. For example, for $B_0=2$ this introduces an order of magnitude approximation error in $\Omega$ and $\Lambda$. Secondly,  there is some fictitious dynamics of the scalar d.o.f., excited by the fact that the BZ parametrization~(\ref{Eq:BZansatz}) does not give an exact representation of the background dynamics. We also expect this discrepancy to be partially due to the fact that the QS approximation inherent in the treatment of $f(R)$ in MGCAMB does not give a full account of the ISW effect. However, in order to make meaningful statements about the latter, we need to make a comparison between the output of MGCAMB and the output of EFTCAMB with the full treatment of the background, i.e. 
consider the designer case mentioned above.

Let us then abandon the BZ parametrization for our background cosmology, and rather adopt the designer approach that allows us to reconstruct all the viable $f(R)$ models that reproduce a $\Lambda$CDM expansion history.
As mentioned earlier, $f(R)$ models are able to reproduce any given expansion history by means of a 
designer approach firstly discussed in~\cite{Song:2006ej} and later generalized to include radiation and 
a time varying dark energy equation of state in~\cite{Pogosian:2007sw}. The Friedmann equation for $f(R)$ theories can indeed be written as a second order differential equation for $f[R(a)]$, namely:
\be\label{fR_designer}
f''-\left( 1+{H'\over H} + {R''\over R'}\right)f'+{R' \over 6H^2}f =- {R' \over 3M_P^2H^2} \rho_{\rm{DE}} ,
\ee
where primes denote differentiation w.r.t. $\ln{a}$ and $\rho_{\rm DE}$ is the energy density of the effective dark energy component. The procedure consists then in fixing the expansion history by choosing  an equation of state of dark energy $w_{\rm DE}(a)$, determining the corresponding energy density like in~(\ref{density_DE}) and solving Eq.~(\ref{fR_designer}) for $f$. For any given expansion history the solution will consist in a family of $f(R)$ models labeled by $B_0$. We implement this procedure in EFTCAMB and show the output in Fig.~\ref{Fig.BackgroundDesigner}; for the $\Lambda$CDM and $w$CDM cases one can notice that the reconstructed $f(R)$ are in agreement with those of~\cite{Song:2006ej}. We show also the results for the case of a CPL background.
\begin{figure}[t!]
\begin{center}
  \includegraphics[width=0.8\textwidth]{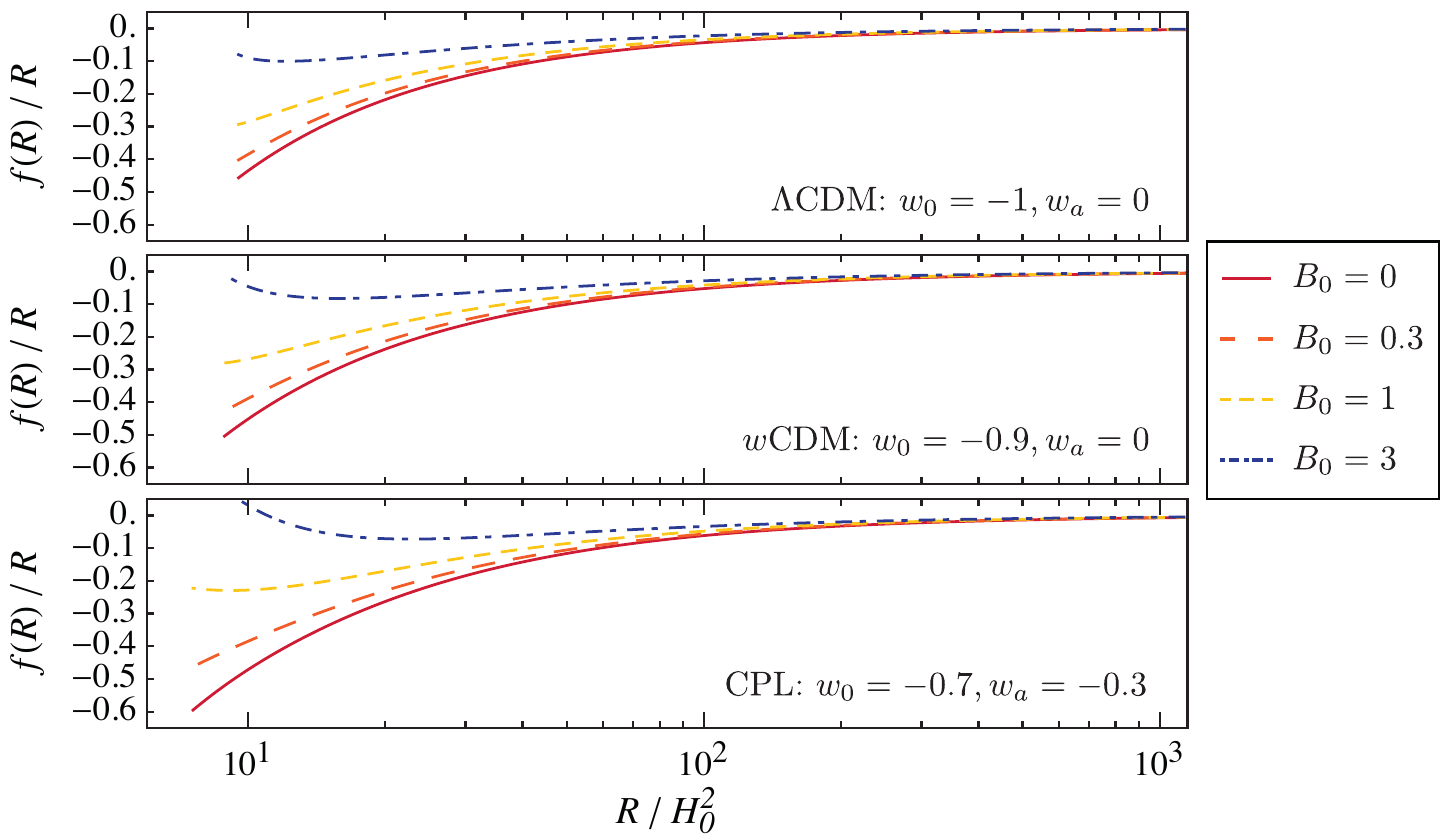}
  \caption{\label{Fig.BackgroundDesigner} We show  designer $f(R)$ models that mimic different expansion histories as reconstructed with our code: $\Lambda$CDM in the top panel, a constant $w_{\rm DE}=-0.9$ in the middle panel and a time-varying $w_{\rm DE}$ in the bottom panel. For each case we plot four curves corresponding to four different values of the boundary condition $B_0$. The values of $w_{\rm DE}$ in the first two panels are chosen to facilitate the comparison with~\cite{Song:2006ej}. See Sec.~\ref{BZ_fR} for a detailed explanation.}
\end{center}
\end{figure}

We start with a $\Lambda$CDM expansion history, consider different values of $B_0$ and compare our results with those of MGCAMB in the lower panel of Fig.~\ref{Fig.Relative_Diff}. The overall agreement for values of $B_0<0.1$ is within $ 0.1\%$ in the high multipoles regime and within $1\%$ in the low multipoles regime. 
For larger values, i.e. $B_0\gtrsim 1$ (which are in tension with constraints from current data~\cite{Lombriser:2010mp,Hu:2013aqa,Marchini:2013oya,Marchini:2013lpp}) we notice that at large scales there is a better agreement, while on smaller scales we get some systematic offset. In what follows we analyze this discrepancy using $B_0=2$ which emphasizes the offset of the codes and facilitates the investigation.
\begin{figure}[t!]
\begin{center}
  \includegraphics[width=0.87\textwidth]{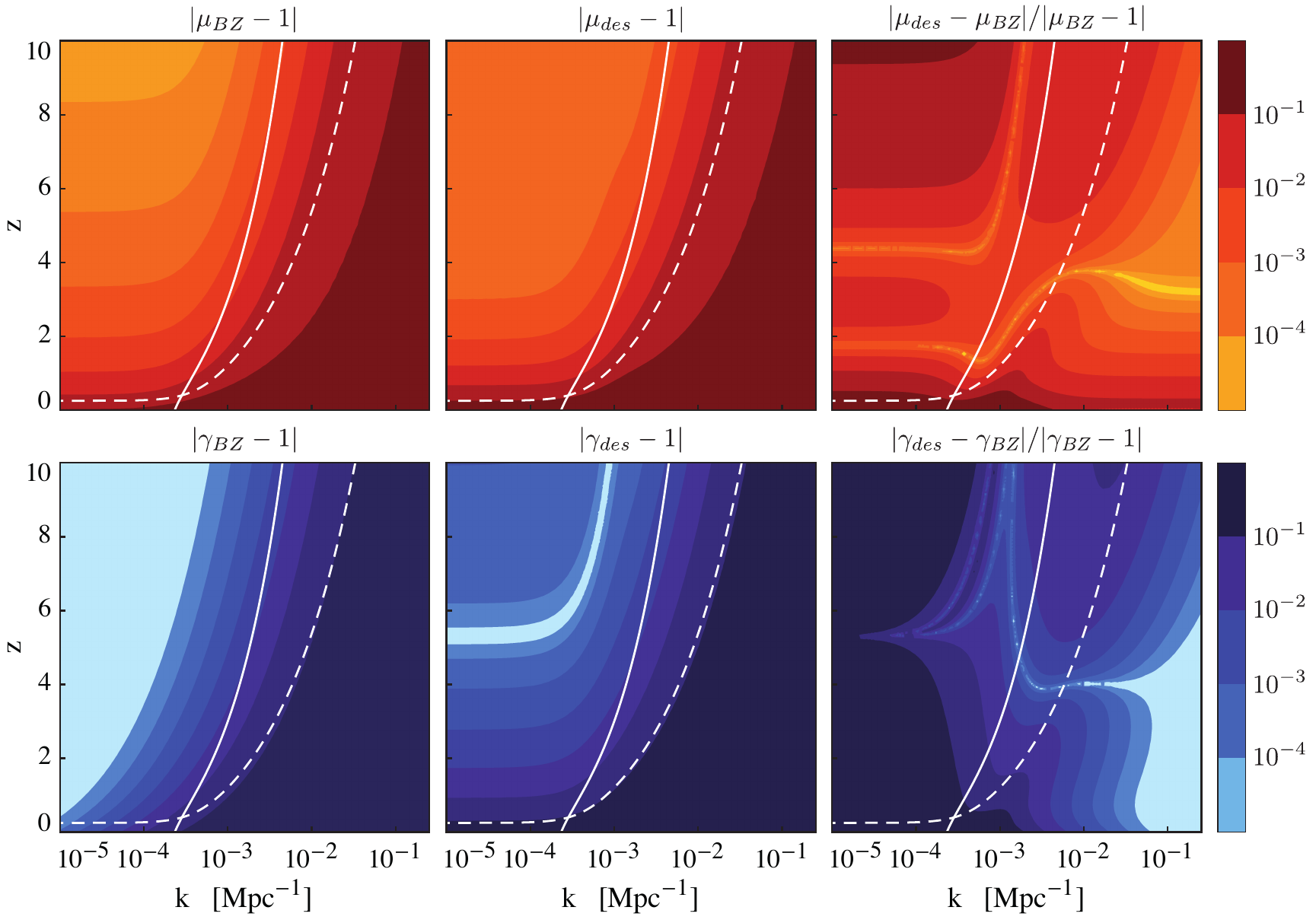}
  \caption{\label{Fig.DesignerMu} We compare the functions $\mu(z,k)$ and $\gamma(z,k)$ in Eq.~(\ref{BZ}), with $B_0=2$ and $s=4$, to those computed in EFTCAMB evolving the full set of Einstein-Boltzmann equations for $f(R)$ models that reproduce a $\Lambda$CDM expansion history and have $B_0=2$. In all plots, the solid  line represents the physical horizon while the dashed line represents the Compton wavelength of the scalaron~(\ref{ComptonWave}).
\emph{Upper panel}: respectively $|\mu_{\rm BZ}-1|$, $|\mu_{\rm des}-1|$ and $|\mu_{\rm BZ}-\mu_{\rm des}|/|\mu_{\rm BZ}-1|$. \emph{Lower panel}: $|\gamma_{\rm BZ}-1|$, $|\gamma_{\rm des}-1|$ and $|\gamma_{\rm BZ}-\gamma_{\rm des}|/|\gamma_{\rm BZ}-1|$. See Sec.~\ref{BZ_fR} for a detailed explanation.}
\end{center}
\end{figure}

We choose to investigate the source of the above mentioned discrepancy by comparing the functions $\mu(z,k)$ and $\gamma(z,k)$ 
in the BZ parametrization~(\ref{BZ}) to those inferred from our code. The latter are obtained evolving the full dynamics 
for the designer $f(R)$ model in EFTCAMB and then substituting the perturbations into Eqs.~(\ref{mugamma}), 
therefore we indicate them with a subscript `des'.  In Fig.~\ref{Fig.DesignerMu} we plot all these quantities in the $(z,k)$ space, 
as well as the fractional difference between the BZ and designer quantities both for $\mu$ and for $\gamma$.  
Overall we get good agreement between the BZ quantities and our designer ones, 
reproducing the known pattern of recovery of the standard GR behavior at early times on large scales, 
and having some significant deviations from the standard behavior on small scales at late times.  
After a more careful look, we see that on super-horizon scales the differences between 
$\mu_{\rm BZ} $($\gamma_{\rm BZ}$) and $\mu_{\rm des}$ ($\gamma_{\rm des}$) are relatively small and are simply due to the fact that our full-Boltzmann code catches some well known dynamics of the scalaron  at those scales, and the return to GR is not as exact as in the quasi-static BZ where it is imposed a priori.  On smaller scales, in particular on scales around the Compton wavelength of the scalaron, the fractional difference plot shows some non-trivial differences between the BZ and designer quantities. In other words, at late times and on scales around the Compton one, EFTCAMB is able to catch some dynamics of the scalaron which is not entirely negligible and perhaps is the source of the discrepancies that we noticed in the CMB lensing spectrum on small scales. The latter appears especially in models for which the Compton wavelength of the scalaron is close to the horizon scale and the sub-horizon and sub-Compton regimes are not clearly distinguished. 

To investigate the non-trivial sub-horizon dynamics further, we introduce the following indicator:
\begin{align}\label{QuasistatInd}
\xi \equiv \frac{\dot{\pi}}{\mathcal{H}\pi} \,,
\end{align}
which quantifies deviations from quasi-staticity for the scalar degree of freedom.  In this context with quasi-staticity we mean the fact that time derivaties of the quantities of interest can be neglected.
\begin{figure}[!]
\begin{center}
  \includegraphics[width=0.9\textwidth]{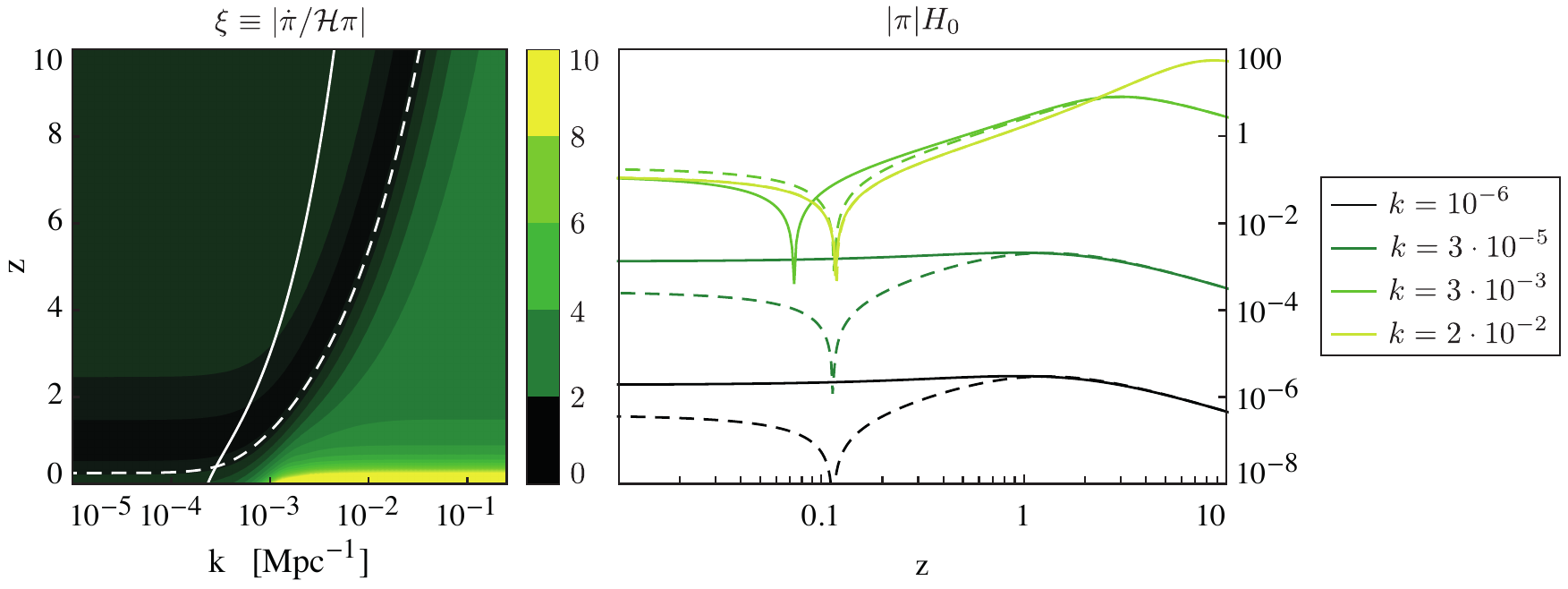}
  \caption{\label{Fig.DesignerQ} \emph{Left:} time and scale dependence of $\xi$ Eq.~(\ref{QuasistatInd}), which is the quantity we introduce as an indicator of the applicability of the QS approximation. \emph{Right:} time evolution of the St$\ddot{\text{u}}$ckelberg field $|\pi| H_0$ (solid curve) compared to that of the source $|H_0E/(C+k^2D)|$~(\ref{pi_inst}) (dashed curve) for four different scales (dark curves are for large scales, light curves are for smaller, but linear, scales). Notice that for $k=2\cdot 10^{-2}$ the source and the field coincide. In both panels we use an $f(R)$ model with a $\Lambda$CDM expansion history and $B_0=2$. See Sec.~\ref{BZ_fR} for a detailed explanation.}
\end{center}
\end{figure} 
We plot $\xi$ in the left panel of Fig.~\ref{Fig.DesignerQ}; from that contour plot one can notice that  the scalaron has some dynamics on super-horizon scales, it then slows on scales of the order of its Compton wavelength and finally resumes evolving in time below the latter scale, especially at low redshift. Let us stress that $\xi$ is a good indicator of whether one can neglect  the time derivatives of the scalar field, but does not necessarily carry information on the dynamics of the metric potentials and therefore on the overall validity of the QS approximation. The latter will depend on how the scalar field couples with gravity and the matter sector.
In the right panel of Fig.~\ref{Fig.DesignerQ} we plot the behavior of $\pi$ as a function of redshift for four different scales, comparing it with the evolution of the source term in Eq.~(\ref{pi_inst}). The curves confirm what we inferred about the dynamics of $\pi$ from the behavior of the indicator $\xi$; on very large scales the scalaron evolves slowly, following the source term at early times and then almost stops evolving at extremely late times. On the other hand, on smaller scales, the field evolves slowly at early times, tracking the source and continues to evolve even at later times eventually crossing zero at some point.
At this point the QS approximation for the dark sector breaks down because the field becomes very small while its derivatives remains finite.

To summarize, we see that the strongest deviations from the BZ parametrization  of $f(R)$ gravity are found close to the Compton wavelength of the scalaron. Below this scale, depending on the value of $B_0$, the dynamics of the scalar field might be non-negligible even if we are on sub-horizon scales; and depending on the coupling of $\pi$ to gravity and dark matter, this might generate a non-standard dynamics of matter perturbations. We expect deviations from QS, parametrized predictions to show up in cosmological observables around $k_c^2 = 6H_0^2/ B_0$ which is roughly the Compton scale today. 
For what concerns the CMB, this  effect will show up, for very large values of $B_0$, both on very large scales due to the differences induced on the ISW effect and on small scales due to the modified evolution of perturbations that will influence the lensing of the CMB.
As the value of $B_0$ decreases the Compton wavelength will move to scales that just contribute to the lensing, but the magnitude of the effect will decrease as well. In the end, for small values of $B_0$, this will just introduce some very small, negligible, discrepancies that we can see in Fig.~\ref{Fig.Relative_Diff}.
We however stress that ongoing experiments, such as \emph{Planck}, and forthcoming ones, like Euclid, are expected to be much more sensitive to these effects which will have to be properly accounted for when extracting predictions for the observables of interest.

\subsubsection{Designer $f(R)$ models on non-$\Lambda$CDM background}\label{designer_fR_2}
\begin{figure}[t!]
\begin{center}
  \includegraphics[width=0.9\textwidth]{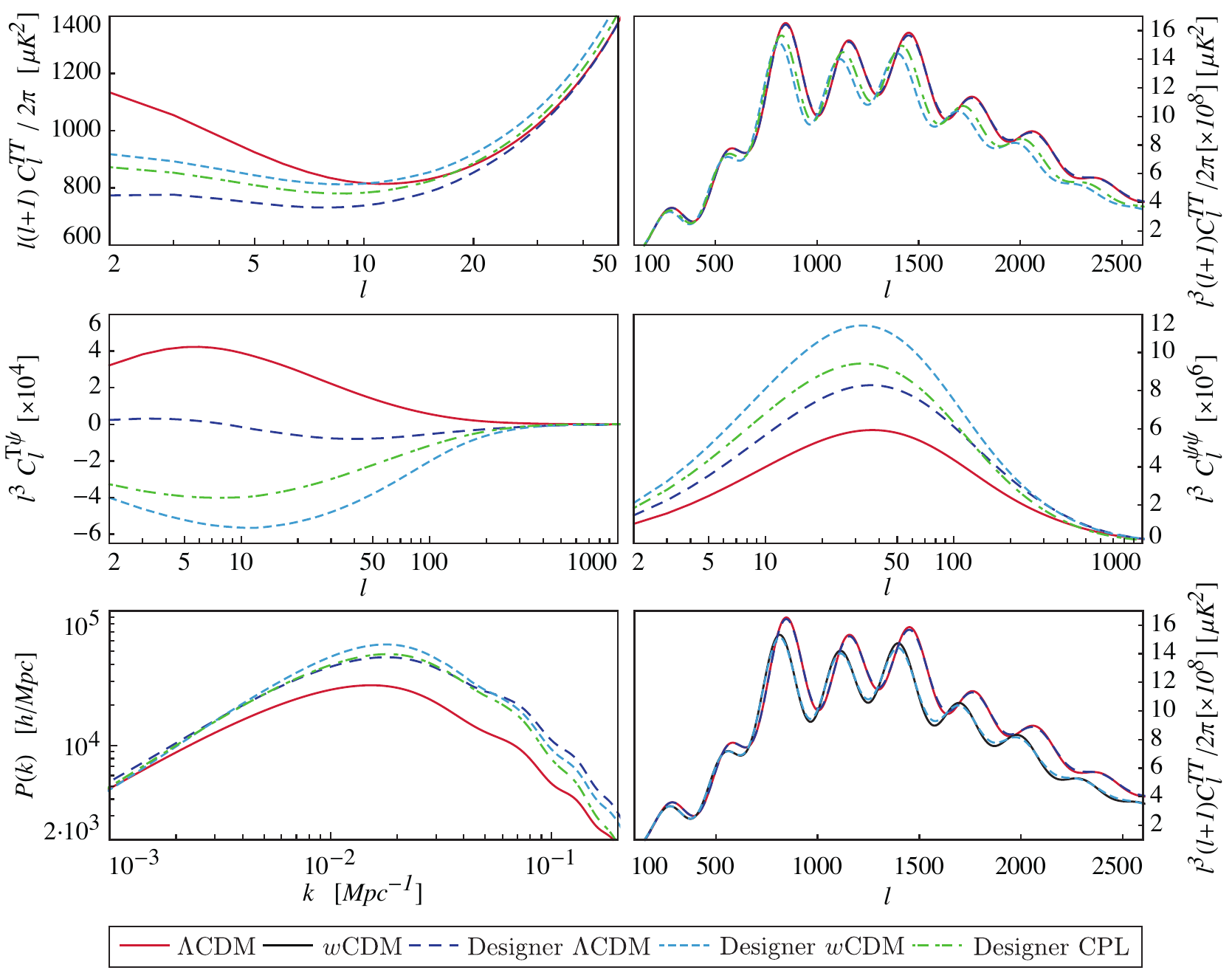}
  \caption{\label{Fig.EFTDesignerSpectra} Power spectra of several cosmological observables for $f(R)$ models mimicking both $\Lambda$CDM and non-$\Lambda$CDM expansion histories. The red solid line represents predictions for the $\Lambda$CDM model while the black solid one stands for a $w$CDM with $w_0=-0.7$ (shown only in the bottom right panel). Dashed lines portray designer $f(R)$ models with different expansion histories but same boundary condition $B_0=1$: the long-dashed dark blue line corresponds to models with a $\Lambda$CDM background, the short-dashed blue to models with a $w$CDM background with $w_0=-0.7$ and the dashed-dotted light blue to models with a CPL background with $w_0=-0.7$ and $w_a=-0.3$.  \emph{Upper panels:} CMB temperature power spectra; \emph{central panels:} lensing-temperature cross-correlation (left) and the lensing potential power spectra (right); \emph{lower panels:} total matter (left) and CMB temperature power spectra (right) for $\
Lambda$CDM / $w$CDM and the corresponding designer $f(R)$ models. See Sec.~\ref{designer_fR_2} for a detailed explanation.}
\end{center}
\end{figure}
In this section we shall use EFTCAMB to compute the power spectra of different cosmological observables for $f(R)$ models that mimic more general expansion histories. As above, after choosing an expansion history, we reconstruct viable models via an implementation in our code of the $f(R)$ designer approach and the matching formulae~(\ref{fR_matching}).
 We consider a $w$CDM expansion history with $w_0=-0.7$, a CPL model with $w_0=-0.7$, $w_a=-0.3$ and we compare the results with those of the $\Lambda$CDM models analyzed in~\ref{BZ_fR}. In all cases we fix $B_0=1$ in order to make the various effects clearly visible. 
In particular we choose the parameters of the CPL model in order to resemble a cosmological constant at high redshift while evolving toward the $w$CDM case at late times. 

We show the power-spectra observables calculated with our code in Fig.~\ref{Fig.EFTDesignerSpectra} and in what follows we give a detailed overview of each result. 
The first, top left, panel shows the ISW part of the CMB temperature power spectrum. On these angular scales we notice the effects of a modified time evolution of the  gravitational potentials at late times, that results in an overall suppression of power at low multipoles. 
This effect will, however, be shaded by cosmic variance which lowers the statistical significance of these deviations. In fact, 
we expect differences from the $\Lambda$CDM behavior at small scale to acquire a primary role in testing alternative models with ongoing and upcoming surveys~\cite{Calabrese:2009tt}.
We zoom in on the modifications to $C_\ell^{TT}$ at small scales  in the top right panel,  where we can more clearly see the part of the temperature power spectrum which is influenced by gravitational lensing. 
As expected, we notice that the change in the expansion history shifts the position of the peaks and reduces their amplitudes,
while the modification of gravity could further smear the acoustic peaks in the lensing part. 
This can be clearly seen in the lower right figure which compares explicitly the resulting temperature spectra from CAMB and our designer EFTCAMB in $\Lambda$CDM and $w$CDM background. 
The impact of modifications of gravity on the CMB lensing potential is shown in the center right figure where we plot $C_\ell^{\psi\psi}$; one can appreciate that the different expansion histories change the angular size of the lenses slightly shifting the position of the peak, while the different dynamics of perturbations greatly impact the amplitude of the spectrum.
Ongoing CMB experiments like \emph{Planck}, ACT and SPT~\cite{Ade:2013tyw,Das:2013zf,vanEngelen:2012va} have directly measured this observable, and in the upcoming future they will measure it with even greater accuracy, so to this extent codes like ours, that evolve the full dynamics and capture interesting features at those scales of the CMB spectrum, will be very useful.

Another quantity which is greatly influenced by modification of gravity is the power spectrum of the cross-correlation between temperature and lensing potential, i.e. $C_\ell^{\psi T}$. 
As we already commented, the evolution of the Weyl potential sources both the ISW and weak lensing effect inducing a correlation between these two. From the center left panel of Fig.~\ref{Fig.EFTDesignerSpectra} one can notice that for the $\Lambda$CDM model the cross-correlation is large and positive, while for $f(R)$ models with a $\Lambda$CDM expansion history but $B_0=1$, the cross-correlation oscillates around zero. Interestingly, the signal can be increased by changing the expansion history while keeping $B_0=1$; in this case the cross-correlation will become large and negative.

Finally, we shall comment on the effects that appear in the total matter power spectrum. In the bottom left panel of Fig.~\ref{Fig.EFTDesignerSpectra} we can appreciate that as soon as $B_0$ is different from zero the spectrum is shifted, both in amplitude and in scale, with respect to the $\Lambda$CDM one.
In addition a non-standard expansion history changes the amplitude of the spectrum at the peak and also the slope at smaller, but still linear, scales as we can see comparing the light blue lines to the dark blue ones.
Interestingly, we can clearly see that  the CPL model lies between the $\Lambda$CDM and the $w$CDM one; the amplitude of the peak, which is influenced by the early time expansion history, lies close to the $\Lambda$CDM one while the slope at smaller scales, which is affected by the late time evolution of matter perturbations, stays close to the $w$CDM model as $w$ is approaching $w_0=-0.7$.
\subsection{Pure EFT parametrizations with phantom-divide crossing}\label{pure_EFT}
\begin{figure}[t!]
\begin{center}
  \includegraphics[width=0.9\textwidth]{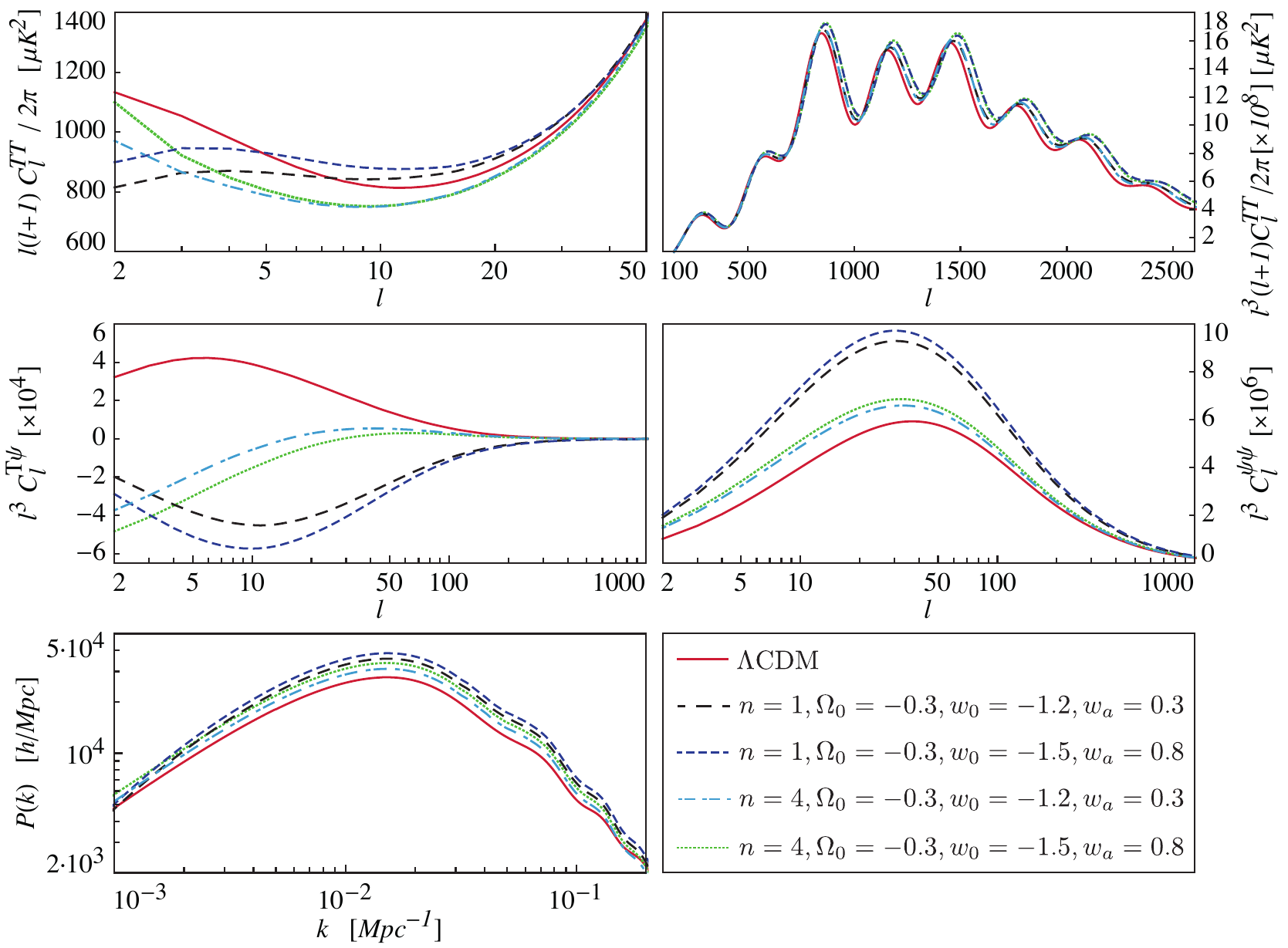}
  \caption{\label{fig:EFTPowerLawModelspectra} Power spectra of several cosmological observables for parametrized EFT models with a phantom-divide crossing background. The red solid line represents predictions for the $\Lambda$CDM model. Dashed lines portray models corresponding to several choices of parameters defining the function $\Omega$. \emph{Upper panels:} CMB temperature power spectra; \emph{central panels:} lensing-temperature cross-correlation (left) and the lensing potential power spectra (right); \emph{lower panels:} total matter power spectra (left). See Sec.~\ref{pure_EFT} for a detailed explanation.}
\end{center}
\end{figure}
In this subsection we shall use our code to explore deviations from $\Lambda$CDM cosmology parametrized with the EFT language.  We will focus on models that contain only background operators.  
On the lines of Sec.~\ref{Sec:background} for the \emph{pure} EFT cases, we fix the desired expansion history and we make a choice for $\Omega(a)$, deriving the remaining quantities from the EFT designer procedure. 

In this paper we consider power laws for $\Omega$, leaving the analysis of the perturbations for other viable choices~\cite{Frusciante:2013zop} for future work.  Specifically we set:
\be
\label{PureEFT:PowLaw}
\Omega(a) = \Omega_0 a^{n} \,,
\ee
which gives an $\Omega$ analogous to the one of the BZ parametrization of $f(R)$ models when  $\Omega_0=-B_0\Omega_m/2$ and $n=3$.
We fix the expansion history to be the one of a dark energy model displaying an extreme phantom-divide crossing which is a phenomenological feature that is naturally and consistently accounted for by the EFT approach. Let us stress that with our code we have checked that these models satisfy the stability constraints listed in Sec.~\ref{Instabilities}. In this case, given that we are not choosing a specific model of DE/MG, but rather a form for $\Omega$, these stability requirements acquire the meaning of a validity check on the time dependence of the EFT functions in view of the corresponding behavior of the perturbations. In particular, the stability conditions will constrain the parameter space describing the expansion history, ($w_0,w_a$), and $\Omega$, in this case $n$, offering a complementary constraining power.

We plot the resulting power spectra in Fig.~\ref{fig:EFTPowerLawModelspectra} for  two ghost-free phantom model: $n=1,4$ and $\Omega_0=-0.3$ in two different background  specified by ($w_0=-1.2,w_a=0.3$) and ($w_0=-1.5,w_a=0.5$).
Aside from the wide array of phenomenological changes that we commented in the previous Section we shall outline here some interesting features.
We can notice that the CMB power spectrum at small scales is mostly influenced by the change in the expansion history while all the other observables are more sensitive to the change in the power law exponent. 
Between the linear ($n=1$) and the non-linear ($n=4$) models we can see a pronounced qualitative difference while the different expansion histories induce just quantitative changes.
This is particularly clear in the ISW part of the CMB temperature power spectrum, in the lensing potential and in the lensing-temperature cross correlation.
Interestingly enough we see that the effects on the total matter power spectrum are limited even if the models that we considered  are chosen to be rather extreme.
At last we notice that no pathological feature arises in these spectra associated to the crossing of the phantom-divide.

\section{Conclusions}\label{Sec:conclusion}

The effective field theory of cosmic acceleration encloses all single field dark energy and modified gravity models within a single parametrization, offering a useful tool to investigate the underlying theory of gravity on large scales in a model-independent way. 
In this effective approach the Lagrangian is constructed following precise and general rules based on the unbroken symmetries of the theory, i.e time-dependent spatial diffeomorphisms. As such, the resulting action contains a finite number of operators at every order in perturbations and derivatives, with each operator multiplied by a free time-dependent function. We referred to the latter as the EFT functions. Three of such operators, equivalently three functions of time, contribute to the background dynamics, while the others affect only the behavior of perturbations. Any model of  DE/MG that introduces a single scalar field and has a well defined Jordan frame (postulated by the validity of the weak equivalence principle) can be mapped precisely into the EFT language via a matching procedure on the EFT functions. 

In this paper we presented an implementation of the EFT framework in the publicly available CAMB code. The resulting product, which we dubbed EFTCAMB, is a full versatile Einstein-Boltzmann code which allows a thorough investigation of  linear scalar cosmological perturbations in general theories that approach the phenomenon of cosmic acceleration. Our code has several advantages. 

Making manifest the multifaceted nature of the EFT formalism, the code we have developed can serve very different purposes. Given the matching of a DE/MG model with  the EFT functions, one can use EFTCAMB to evolve the full dynamics of scalar perturbation on all linear scales in the chosen model. In this paper we gave an example of this by studying the dynamics of perturbations in viable $f(R)$ models with different expansion histories. In the future we plan to implement a built-in matching to popular models of dark energy and modified gravity. Depending on the tests one is interested in performing, EFTCAMB can be used also to implement parametrized EFT models, where one does not specify any theory, but rather chooses different ans$\ddot{\text{a}}$tze for the EFT functions and explore their impact on cosmological observables. In this paper we have shown some examples with  parametrized models involving only the EFT background functions. However, our setup is more general than that and it includes 
contributions from all second order operators. We leave  a thorough investigation of the cosmological implications of all EFT operators affecting linear perturbations for future work.  

EFTCAMB does not rely on any QS approximation, solving instead the full dynamics of  scalar perturbations on all linear scales. The latter is an important feature for several reasons; it allows the exact implementation of  any given single field DE/MG model that can be cast into the EFT language, without need of resorting to sub-horizon approximated expressions which often correspond to solutions of the theory. It also ensures that we do not miss out on any potential sub-horizon dynamics of the scalar d.o.f. which in some cases could be non-negligible and could leave an imprint on cosmological observables within the reach of ongoing and upcoming surveys. To this extent, we presented an example of signature on the small scales of the CMB lensing potential angular power spectrum from the sub-horizon dynamics of the scalaron in $f(R)$ models. Effects like the latter will be measured at increasing accuracy in the next years,  and our code offers a way to exploit these data as complementary tests of gravity.  

Our code can handle very general expansion histories, ranging from the $\Lambda$CDM one to phantom-divide crossing ones. In this paper we presented results for models with a $\Lambda$CDM, $w$CDM and CPL background (including phantom-divide crossing models). The time-varying case is implemented in such a way that the contribution to perturbations coming from the evolution of the dark energy equation of state, is consistently taken into account. Finally, our code has a built-in check for theoretical viability of the model under consideration. In other words, in order to ensure that the underlying theory of gravity is physically acceptable we impose on the EFT operators some stability conditions and the code automatically checks that they are satisfied before proceeding with the evolution of the equations. In particular we require: a positive Newtonian constant, absence of ghost instabilities, absence of super-luminary propagating perturbations and, finally, a positive mass of the extra scalar degree of freedom.

Let us briefly recap the numerical results presented in this paper. We started comparing our code with the outputs of MGCAMB for $f(R)$ theories in a $\Lambda$CDM background. To this extent, we focused on the CMB temperature angular spectrum and showed an agreement of the two code within $0.1\%$ for values of the  
scalaron Compton wavelength consistent with existing bounds from \emph{Planck}. For larger values of $B_0$ we found some tension both at the low and high multipoles; the former is due to the fact that not relying on any QS approximation, EFTCAMB gives a more accurate account of the ISW effect. To investigate the latter discrepancy instead, we considered the functions $\mu(z,k)$ and $\gamma(z,k)$ Eqs.~(\ref{mugamma}), commonly used to parametrize deviations from GR  and implemented in MGCAMB. We compared their shape in $(z,k)$-space as reconstructed from the full evolution of the perturbations in our code, to their form under the BZ approximation~(\ref{BZ}) employed to study $f(R)$ in MGCAMB. This comparison showed that EFTCAMB catches the mild dynamics of the scalaron at early times and on large scales, as well as some non-trivial dynamics on scales around and below the Compton wavelength. We confirmed this by analyzing the time- and scale-dependence of a quantity that we propose as an indicator of quasi-staticity in the dark sector, $\xi$ Eq.~(\ref{QuasistatInd}). 
After this thorough check of consistency with MGCAMB, we moved on to fully exploit the flexibility of the EFT framework applying our code to two different dark energy scenarios. 
First, as an example of what we called the \emph{mapping} EFT approach, we extended the implementation of the $f(R)$ designer approach to more general expansion histories like the $w$CDM and CPL one, examining  the effects of the combined change in the background dynamics and in the growth of structure on cosmological observables like the CMB temperature and lensing power spectra (auto- and cross-correlation), and matter power spectrum.
Second, we used the \emph{pure} EFT approach, i.e. we chose an expansion history and a parametrized form for the EFT function $\Omega$ and again explored signatures on power-spectra observables. In this case we focused on backgrounds with  a phantom-divide crossing demonstrating how within the EFT framework there is no special pathology arising when $w_{\rm DE}=-1$ is crossed.

Let us conclude with some remarks about future directions. We envisage several applications of our code. To start with, we plan to build-in into the code  the matching to several dark energy and modified gravity models and perform parameter space exploration on some of these models. We will perform an analysis of viable parametrized EFT models based also on the findings of~\cite{Frusciante:2013zop}, as well as a thorough investigation of the validity of the QS approximation on sub-horizon scales for general models of dark energy. We also plan to make the code publicly available in the near future. We believe that EFTCAMB will be a very useful tool to explore the full dynamics of perturbations  in single scalar field approaches to the phenomenon of cosmic acceleration, providing upcoming surveys such as Euclid with a powerful code to perform both model-dependent and -independent tests of gravity on large scales.

\acknowledgments
We are grateful to Carlo Baccigalupi, Jolyon Bloomfield, Paolo Creminelli, Stefano Liberati, Federico Piazza, Levon Pogosian and Riccardo Valdarnini for useful conversations. 
BH is indebted to Nicola Bartolo and Sabino Matarrese for various discussions at the initial stage of this work. AS is grateful to Alireza Hojjati, Levon Pogosian and Gong-Bo Zhao for their previous and ongoing collaboration on related topics.
BH is supported by the Dutch Foundation for Fundamental Research on Matter (FOM).
NF acknowledges partial financial support from the European Research Council under the European Union's Seventh 
Framework Programme (FP7/2007-2013) / ERC Grant Agreement n.~306425 ``Challenging General Relativity'' and from 
the Marie Curie Career Integration Grant LIMITSOFGR-2011-TPS Grant Agreement n.~303537. 
AS acknowledges support from a SISSA Excellence Grant. MR and AS acknowledge partial support from the INFN-INDARK initiative.

\appendix

\section{Contributions from second order operators}\label{higher_orders}
The effective field theory of dark energy action in conformal time with the $\pi$ field  manifest through the St$\ddot{\text{u}}$ckelberg trick, up to second order operators, read:
\begin{align}\label{full_action_Stuck}
S = \int d^4x \sqrt{-g}& \left \{ \frac{m_0^2}{2} \l[1+\Omega(\tau+\pi)\r]R+ \Lambda(\tau+\pi) - c(\tau+\pi)a^2\left[ \delta g^{00}-2\frac{\dot{\pi}}{a^2} + 2\hub\pi\left(\delta g^{00}-\frac{1}{a^2}-2\frac{\dot{\pi}}{a^2}\right) +2\dot{\pi}\delta g^{00} \right.\right. \nonumber \\
 &\left.\left.+2g^{0i}\partial_i\pi-\frac{\dot{\pi}^2}{a^2}+ g^{ij}\partial_i \pi \partial_j\pi -\l(2\hub^2+\dot{\hub}\r)\frac{\pi^2}{a^2} + ... \right] \right. \nonumber \\
 &\left. + \frac{M_2^4 (\tau + \pi)}{2}a^4 \left(\delta g^{00} - 2 \frac{\dot{\pi}}{a^2}-2\frac{\hub\pi}{a^2}+... \right)^2 \right. \nonumber\\
& \left.- \frac{\bar{M}_1^3 (\tau + \pi)}{2}a^2 \left(\delta g^{00}- 2 \frac{\dot{\pi}}{a^2}-2\frac{\hub\pi}{a^2}+... \right) \left(\delta 
\tensor{K}{^\mu_\mu} + 3 \frac{\dot{\hub}}{a} \pi + \frac{\SDer^2 \pi}{a^2}+ ...\right) \right. \nonumber \\
&\left. - \frac{\bar{M}_2^2 (\tau + \pi)}{2} \left(\delta \tensor{K}{^\mu_\mu} + 3 \frac{\dot{\hub}}{a} \pi + \frac{\bar{\nabla}^2 \pi}{a^2} + ... \right)^2 \right.\nonumber  \\
& \left. - \frac{\bar{M}_3^2 (\tau + \pi)}{2}
  \left(\delta \tensor{K}{^i_j} + \frac{\dot{\hub}}{a} \pi \delta\indices{^i_j}
  + \frac{1}{a^2} \bar{\nabla}^i \bar{\nabla}_j \pi +... \right)
  \left(\delta \tensor{K}{^j_i} + \frac{\dot{\hub}}{a} \pi \delta\indices{^j_i}
  + \frac{1}{a^2} \SDer^j \SDer_i \pi + ... \right) \nonumber \right. \\
&  \left. + \frac{\hat{M}^2 (\tau + \pi)}{2} a^2 \left(\delta g^{00} - 2 \frac{\dot{\pi}}{a^2} - 2\frac{\hub}{a^2}\pi + ... \right)\, \left(\delta R^{(3)} +4\frac{\hub}{a} \bar{\nabla}^2\pi + ...\right)\right. \nonumber \\
&\left. + m_2^2(\tau+\pi)\left(g^{\mu\nu}+n^{\mu} n^{\nu}\right)\partial_{\mu}\left(a^2g^{00}-2\dot{\pi} -2\hub\pi + ...\right)\partial_{\nu}\left(a^2g^{00}-2\dot{\pi} - 2\hub\pi + ...\right) + ...\right\}  + S_{m} [g_{\mu \nu}],
\end{align}
where $\bar{\nabla}$ indicates three dimensional spatial derivatives. Note that the conformal scale factor has been already Taylor expanded in $\pi$  according to Eq.~\ref{taylorexpansionstuck}.

In what follows we list the contributions to the linearly perturbed equations of Sec.~\ref{Sec:lin_perturbations} from the second order operators in~(\ref{full_action_Stuck}).  Let us make an itemized list where for each operator we list its contributions to the r.h.s. of Eq.~(\ref{00}) by $\Delta_{00}$, to the r.h.s. of Eq.~(\ref{0i}) by $\Delta_{0i}$, to the r.h.s. of Eq.~(\ref{ijoff}) by $\Delta_{ij,i\neq j}$, to the r.h.s of Eq.~(\ref{space-space_trace}) by $\Delta_{ii}$ and to the l.h.s. of Eq.~(\ref{pi0}) by $\Delta_{\pi}$. Notice that in order to perform a correct stability analysis on the equation for $\pi$, along the lines of Sec.~\ref{Instabilities}, it is important to demix the degrees of freedom; specifically, \emph{once the contributions from all operators have been taken into account}, one needs to use the Einstein equations to subsistute for any $\ddot{h}, \eta, \sigma_*$ appearing in the final form of the equation for $\pi$.

\vspace{1cm}
\begin{itemize}
\item[] \underline{$(\delta g^{00})^2$}:\\
{\small
\ba
\label{Delta_a_00}
&&\Delta_{00}=-\f{2M_2^4a^2}{m_0^2(1+\Omega)}\l(\dot{\pi}+\hub\pi\r) \,,\nonumber  \\
\label{Delta_a_0i}
&&\Delta_{0i}=0 \,,\nonumber \\
\label{Delta_a_ijoff}
&&\Delta_{ij,i\neq j}=0 \,,\nonumber\\
\label{Delta_a_ii}
&&\Delta_{ii}= 0 \,,\nonumber\\
\label{Delta_a_pi}
&&\Delta_{\pi}=2M_2^4\l[\ddot{\pi}+4\l(\hub+\f{\dot{M}_2}{M_2}\r)\dot{\pi}+\l(3\hub^2+\dot{\hub}+4\f{\hub\dot{M}_2}{M_2}\r)\pi\r] \,.  
\ea}
\item[] 
\underline{$\delta g^{00}\delta K^\mu_\mu$}:\\
{\small
\ba
\label{Delta_b_00}
&&\Delta_{00}=\f{a\bar{M}_1^3}{2m_0^2(1+\Omega)}\l[k{\mathcal Z}-3\l(\dot{\hub}-2\hub^2-\f{k^2}{3}\r)\pi+3\hub\dot{\pi}\r] \,,\nonumber \\
\label{Delta_b_0i}
&&\Delta_{0i}= \f{a\bar{M}_1^3k}{m_0^2(1+\Omega)}\l(\dot{\pi}+\hub\pi\r) \,,\nonumber\\
\label{Delta_b_ijoff}
&&\Delta_{ij,i\neq j}=0 \,,\nonumber\\
\label{Delta_b_ii}
&&\Delta_{ii}=-\f{3a\bar{M}_1^3}{m_0^2(1+\Omega)}\l[\ddot{\pi}+\l(4\hub+3\f{\dot{\bar{M}}_1}{\bar{M}_1}\r)\dot{\pi}+\l(3\hub^2+\dot{\hub}+3\f{\hub\dot{\bar{M}}_1}{\bar{M}_1}\r)\pi\r] \,,\nonumber\\
\label{Delta_b_pi}
&&\Delta_{\pi}=\f{\bar{M}_1^3}{2a}\l[\l(3\hub+3\f{\dot{\bar{M}}_1}{\bar{M}_1}\r)\l(-k\zed+3(\dot{\hub}-\hub^2)\pi-k^2\pi\r)-\f{\ddot{h}}{2}+\hub k\zed+2\hub k^2\pi+3(\ddot{\hub}-4\hub\dot{\hub}+2\hub^3)\pi \r] \,.
\ea}
\item[] \underline{$(\delta K)^2$}:\\
{\small
\ba
\label{Delta_c_00}
&&\Delta_{00}= \f{3\hub \bar{M}_2^2}{2m_0^2(1+\Omega)}\l[k\zed-3(\dot{\hub}-\hub^2)\pi+k^2\pi\r] \,, \nonumber  \\
\label{Delta_c_0i}
&&\Delta_{0i}= \f{\bar{M}_2^2k}{m_0^2(1+\Omega)}\l[k\zed-3(\dot{\hub}-\hub^2)\pi+k^2\pi\r] \,, \nonumber \\
\label{Delta_c_ijoff}
&&\Delta_{ij,i\neq j}=0 \,,\nonumber\\
\label{Delta_c_ii}
&&\Delta_{ii}=-\f{3\bar{M}_2^2}{m_0^2(1+\Omega)}\l(2\hub+\partial_\tau+2\f{\dot{\bar{M}}_2}{\bar{M}_2}\r)\l[k\zed-3(\dot{\hub}-\hub^2)\pi+k^2\pi\r] \,,\nonumber\\
\label{Delta_c_pi}
&&\Delta_{\pi}=\f{\bar{M}_2^2}{2a^2}\l(3(\dot{\hub}-\hub^2)-k^2\r)\l[-k\zed+3(\dot{\hub}-\hub^2)\pi-k^2\pi\r] \,.
\ea}
\item[] \underline{$\delta K^\mu_\nu\delta K^\nu_\mu$}:\\
{\small
\ba
\label{Delta_d_00}
&&\Delta_{00}= \f{\hub\bar{M}^2_3}{m_0^2(1+\Omega)} \l[k\zed-3\l(\dot{\hub}-\hub^2-\f{k^2}{3}\r)\pi\r] \,,\nonumber  \\
\label{Delta_d_0i}
&&\Delta_{0i}= \f{\bar{M}^2_3k}{m_0^2(1+\Omega)}\l[\f{k\zed}{3}+\f{2}{3}k\sigma_*-(\dot{\hub}-\hub^2-k^2)\pi\r] \,, \nonumber
\ea
\ba
\label{Delta_d_ijoff}
&&\Delta_{ij,i\neq j}=\f{\bar{M}^2_3}{m_0^2(1+\Omega)}\l(2\hub+2\f{\dot{\bar{M}}_3}{\bar{M}_3}+\partial_\tau\r)\l(k\sigma_*+k^2\pi\r) \,,\nonumber\\
\label{Delta_d_ii}
&&\Delta_{ii}=\f{\bar{M}^2_3}{m_0^2(1+\Omega)}\l(2\hub+2\f{\dot{\bar{M}}_3}{\bar{M}_3}+\partial_\tau\r)\l[-k\zed+3\l(\dot{\hub}-\hub^2-\f{1}{3}k^2\r)\pi\r] \,,\nonumber\\
\label{Delta_d_pi}
&&\Delta_{\pi}=\f{\bar{M}^2_3}{a^2}\l[\l(\f{k^4}{2}-k^2(\dot{\hub}-\hub^2)+\f{3}{2}(\dot{\hub}-\hub^2)^2\r)\pi+\l(\f{k^2}{2}-\f{\dot{\hub}-\hub^2}{2}\r)k\zed +\f{k^3}{3}(\sigma_*-\zed)\r] \,.
\ea}
\item[] \underline{$\delta g^{00}\delta R^{(3)}$}:\\
{\small
\ba
\label{Delta_e_00}
&&\Delta_{00}= -\f{2\hat{M}^2}{m_0^2(1+\Omega)}k^2\l(\eta+\hub\pi\r) \,, \nonumber  \\
\label{Delta_e_0i}
&&\Delta_{0i}= 0 \,,\nonumber \\
\label{Delta_e_ijoff}
&&\Delta_{ij,i\neq j}=2\f{\hat{M}^2}{m_0^2(1+\Omega)}k^2\l(\dot{\pi}+\hub\pi\r) \,,\nonumber\\
\label{Delta_e_ii}
&&\Delta_{ii}=\f{4\hat{M}^2}{m_0^2(1+\Omega)}k^2\l(\dot{\pi}+\hub\pi\r) \,,\nonumber\\
\label{Delta_e_pi}
&&\Delta_{\pi}= \f{2k^2}{a^2}\l[\hat{M}^2\f{k}{3}(\sigma_*-\zed)+\l(\hub\hat{M}^2+2\hat{M}\dot{\hat{M}}\r)\eta+\l(2\hub\hat{M}\dot{\hat{M}}+\dot{\hub}\hat{M}^2\r)\pi\r]\,.
\ea}
\item[] \underline{ $\rm(g^{\mu\nu}+n^\mu n^\nu)\partial_\mu\delta g^{00}\partial_\nu \delta g^{00}$}:\\
{\small
\ba
\label{Delta_f_00}
&&\Delta_{00}= -\f{4m_2^2}{m_0^2(1+\Omega)}k^2\l(\dot{\pi}+\hub\pi\r) \,,\nonumber\\
\label{Delta_f_0i}
&&\Delta_{0i}=0  \,, \nonumber\\
\label{Delta_f_ijoff}
&&\Delta_{ij,i\neq j}=0 \,,\nonumber\\
\label{Delta_f_ii}
&&\Delta_{ii}=0 \,,\nonumber\\
\label{Delta_f_pi}
&&\Delta_{\pi}= 4m_2^2\l(2\hub+\f{\dot{m}_2}{m_2}+\partial_\tau\r)k^2\l(\dot{\pi}+\hub\pi\r)\,. 
\ea}
\end{itemize}


\end{document}